\documentclass[fleqn,10pt]{wlscirep}
\usepackage[utf8]{inputenc}
\usepackage[T1]{fontenc}
\usepackage{hyperref}
\usepackage{ulem}

\title{Scaling theory of fractal complex networks}

\author[*]{Agata Fronczak}
\author{Piotr Fronczak}
\author{Mateusz Samsel}
\author{Kordian Makulski}
\author{Michał \L epek}
\author{Maciej J. Mrowinski}
\affil{Faculty of Physics, Warsaw University of Technology, Koszykowa 75, PL-00-662 Warsaw, Poland}
\affil[*]{agata.fronczak@pw.edu.pl}

\begin{abstract}
	We show that fractality in complex networks arises from the geometric self-similarity of their built-in hierarchical community-like structure, which is mathematically described by the scale-invariant equation for the masses of the boxes with which we cover the network when determining its box dimension. This approach - grounded in both scaling theory of phase transitions and renormalization group theory - leads to the consistent scaling theory of fractal complex networks, which complements the collection of scaling exponents with several new ones and reveals various relationships between them. We propose the introduction of two classes of exponents: microscopic and macroscopic, characterizing the local structure of fractal complex networks and their global properties, respectively. Interestingly, exponents from both classes are related to each other and only a few of them (three out of seven) are independent, thus bridging the local self-similarity and global scale-invariance in fractal networks. We successfully verify our findings in real networks situated in various fields (information – the World Wide Web, biological – the human brain, and social – scientific collaboration networks) and in several fractal network models. 
\end{abstract}
\begin{document}
	
	\flushbottom
	\maketitle
	\thispagestyle{empty}
	
\section*{Revising paradigms of fractal complex networks}
	
It will soon be two decades since it was first shown that some real networks (such as the World Wide Web [WWW] and different biological networks) have fractal properties \cite{2005SongNature, 2006SongNatPhys}. This means that, when covered with non-overlapping boxes, with the maximum distance between any two nodes in each box less than $l_B$, they exhibit power-law scaling \cite{2005SongNature, 2006SongNatPhys, 2009RozenfeldEncyclopedia, 2020bookRosenberg, 2021WenFusion}: 
\begin{equation}\label{NB}
	N_B(l_B)/N\simeq l_B^{-d_{B}},
\end{equation}
where $N_{B}(l_B)$ is the number of boxes of a given diameter, and $d_B$ is the fractal (or box) dimension of the network of size $N$. Such fractal networks are also said to be \textit{self-similar}, because their power-law degree distributions, 
\begin{equation}\label{Pk}
	P(k)\sim k^{-\gamma},
\end{equation}	
remain \textit{invariant} under a renormalization scheme \cite{2008RadicchiPRL, 2010RozenfeldPRL}, according to which a new network emerges from the original one when nodes belonging to the same box in the original network are replaced by one supernode in the renormalized network. In this case, the supernode is connected to another supernode if in the original network there is at least one link between the nodes of the corresponding boxes. 

Here, at least two critical remarks can be made. The first remark is that an analogous invariance of the degree distribution with respect to the box-covering renormalization scheme is also observed in networks that do not satisfy Eq.~(\ref{NB}) (in this respect, well-known examples are the internet and Barab\'asi-Albert (BA) networks \cite{2006SongNatPhys, 2007KimNewJPhys, 2008KimJKorean}). The second remark is that it is not entirely clear, what structural characteristics of fractal networks exhibits geometric self-similarity and remain invariant \cite{1997bookDubrulle} under the described renormalization. Clearly, the power-law node degree distribution cannot be considered such a characteristic because it is intrinsically invariant under the rescaling of the degree \cite{2006bookSornette}. Its invariance under box-covering renormalization may only suggest the existence of some (presumably) degree-dependent network measure, whose self-similarity under the renormalization procedure could result in the observed invariance of the degree distribution. One argument supporting this statement is that random networks, where the degree distribution is not a power law, can also exhibit fractal properties (in this regard, the best example is the giant component of classical random graphs near the percolation transition).

If the above remarks, indicating an incomplete understanding of fractality in complex networks, are reasonable, pertinent questions would be: What are the real origins and potential consequences of fractality in complex networks? What determines networks' fractal dimension? Indeed, several studies have been published throughout the years that focus on the exploration of the origins of fractality \cite{2005YookPRE, 2006GohPRL, 2007KimPRE, 2007KitsakPRE, 2008GallosPRL, 2016WeiPRE, 2017FujikiEPJB, 2023ZakarNetSci}. However, these efforts did not lead to consensus. Thus, there is a lack of realistic (and not just deterministic \cite{2007RozenfeldNewJPhys, 2022YakuboPLOS}, or reflecting the renormalization procedure \cite{2006SongNatPhys, 2015KuangSciChina, 2022MolontayInProc}) fractal network models that would allow testing the role of fractality in the context of geometry-involving issues \cite{2021BogunaNatPhys}, such as navigability, localization of information sources, prediction of hidden network connections, etc. These are of particular importance when faced with the confirmed fractal properties of different information, biological, and even social networks (see e.g.\cite{2007GallosPNAS, 2013GallosPLoS}). The goal of this article is to initiate far-reaching changes in this state of affairs.

In what follows, we will first argue that the correct scale-dependent network measure, which is self-similar (i.e. geometrically invariant) under the $l_B$-box-covering renormalization procedure, is the normalized mass of the box - $\mu(L,k)=m(L,k)/\!\langle m\rangle$, where $m(L,k)$ is the number of nodes in the box of diameter $L\geq l_B$ and hub degree $k$, and $\langle m\rangle=N/N_B(L)=L^{d_B}$ (\ref{NB}) is the average mass of non-overlapping boxes of this diameter. It should be emphasized here that although the definition of the box is the same throughout the paper, we distinguish between $l_B$-boxes used to renormalize the network and $L$-boxes (where $L\geq l_B$) whose self-similarity we examine. This distinction is crucial to make it easier to understand the main idea of the paper.

Then, we show that one of the consequences of this result is the previously discovered scaling relation between the degree $k'$ of the supernode in the renormalized network and the degree $k$ of the hub of the corresponding $l_B$-box in the network before renormalization: $k'=l_B^{-d_k}k$, where $d_k$ is only one of four scaling exponents that characterize microscopic structure of the fractal complex network and determine its box dimension. We also show that if the fractal complex network has a power-law node degree distribution (which is traditionally referred to as the \textit{scale-free property}), then the mass box distribution also follows the power-law, and it is invariant under the box renormalization procedure. Furthermore, the characteristic exponents of both distributions are related to the microscopic scaling exponents describing the masses of the boxes, thus bridging local self-similarity and global scale invariance in fractal complex networks. Lastly, we successfully verify our findings in real networks situated in various fields (information – the World Wide Web, biological – the human brain, and social – scientific collaboration networks) and in several fractal network models.

\section*{Local self-similarity and global scale-invariance in fractal networks}

\subsection*{Geometric self-similarity}

In classical fractals \cite{1988bookFeder}, which reproduce themselves at different space scales, self-similarity manifests itself in the scale-invariant equation \cite{1997bookDubrulle}, which describes how the mass $m(L)$ of the system changes with its linear size $L$:
\begin{equation}\label{intro1}
	m(bL)=\mu(b)\,m(L),
\end{equation} 
where $b>0$. In theoretical physics, this type of equation is, for example, encountered in the theory of critical phenomena \cite{2006bookSornette, 2004bookMcComb}. Mathematically, this equation defines a homogeneous function. Its solution is simply a power law:
\begin{equation}\label{intro2}
	m(L)=AL^{d_f},
\end{equation}
which, in the case of fractals, determines their fractal dimension, $d_f=\ln\mu/\ln b$, and leads to the well-known scaling relation \cite{2009BundeEncyclopedia}: 
\begin{equation}\label{intro3}
	m(bL)=b^{d_f}m(L).
\end{equation}

Moving forward, to address the problem of geometric self-similarity in complex networks, we first argue that Eq.~(\ref{NB}) can be treated as a special case of Eq.~(\ref{intro3}). Then, building on this observation, we assume that Eq.~(\ref{intro3}) is also a special case of a more general equation in which the masses of the system and its parts, which are further identified with the number of nodes in the network and the number of nodes in different $L$-boxes extracted from this network, respectively, do not only depend on the diameter of the examined set of nodes (i.e. the entire network or a box) but also on the degree of the best-connected node in this set. This assumption leads us to the consistent scaling theory of fractal complex networks.

\begin{figure}[t]
	\centering
	\includegraphics[width=0.9\columnwidth]{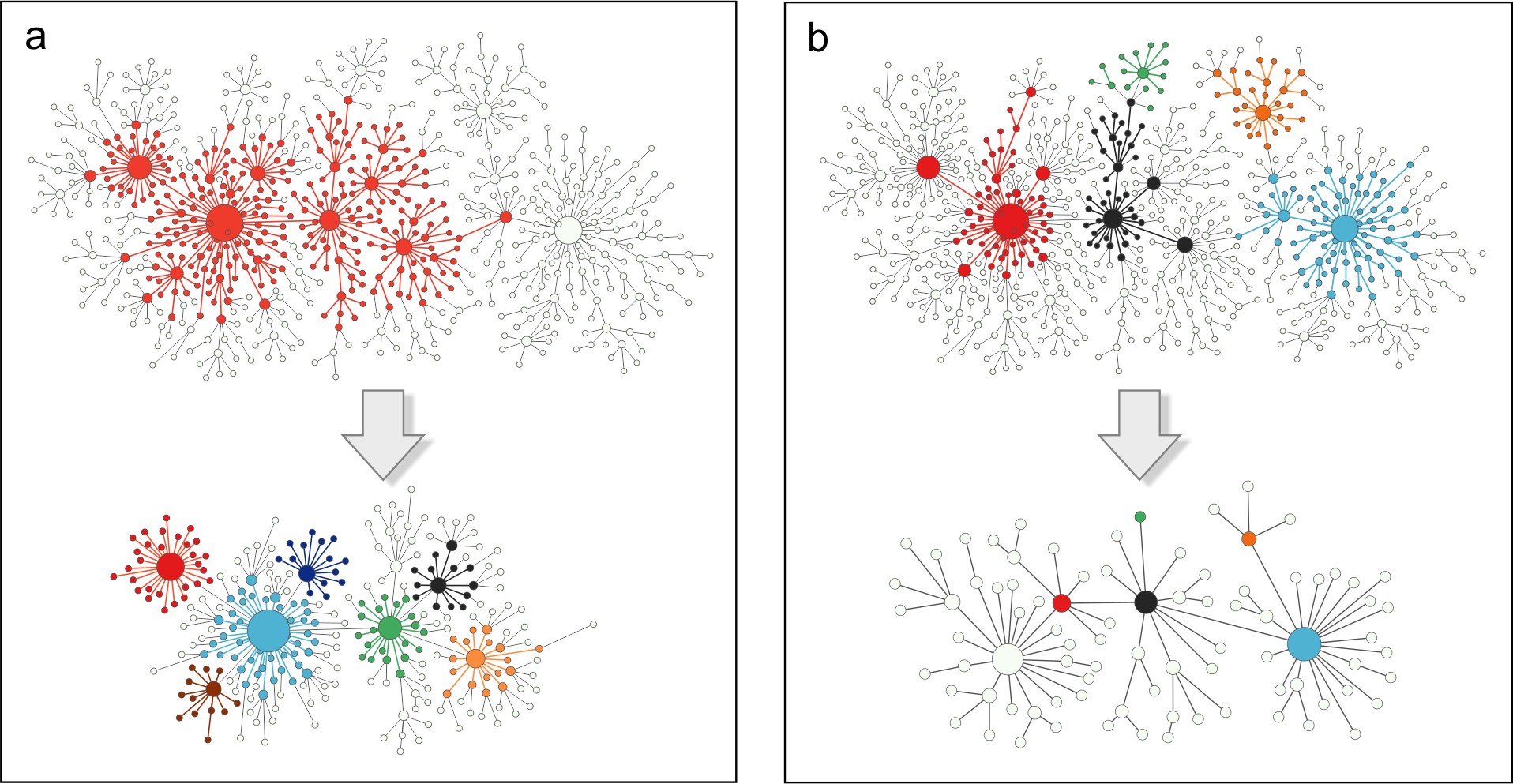}
	\caption{\textbf{Schematic illustration of the idea of geometric self-similarity in complex networks on the example of the fractal model of nested BA networks} (for the definition of the model, see \hyperref[SecMethods]{“Methods”} section). Part \textbf{a)} of the figure shows that the network can be subdivided into parts---boxes of a given diameter---each of which is (at least approximately) a reduced-size copy of the entire network. In the top picture shown, one such box, marked in red, is extracted from the original network and treated as a new network (shown below). It is divided again into new smaller boxes, some of which are marked with different colours. Both macroscopic and microscopic characteristics of this new network (represented by green squares in Fig.~\ref{figmain2}) are similar to those of the original network (indicated by navy circles in Fig.~\ref{figmain2}). Part \textbf{b)} of this figure illustrates renormalization procedure applied to the same network as in part a. The top original network is divided into boxes of a fixed diameter, some of which are marked with different colours. In the new network after renormalization (shown below), these boxes are replaced by nodes with the corresponding colours. Again, the macroscopic and microscopic characteristics of the network after renormalization (represented by red triangles in Fig.~\ref{figmain2}) are similar to those of the original network.\label{figmain1}}
\end{figure}

\begin{figure}[t]
	\centering
	\includegraphics[width=0.75\columnwidth]{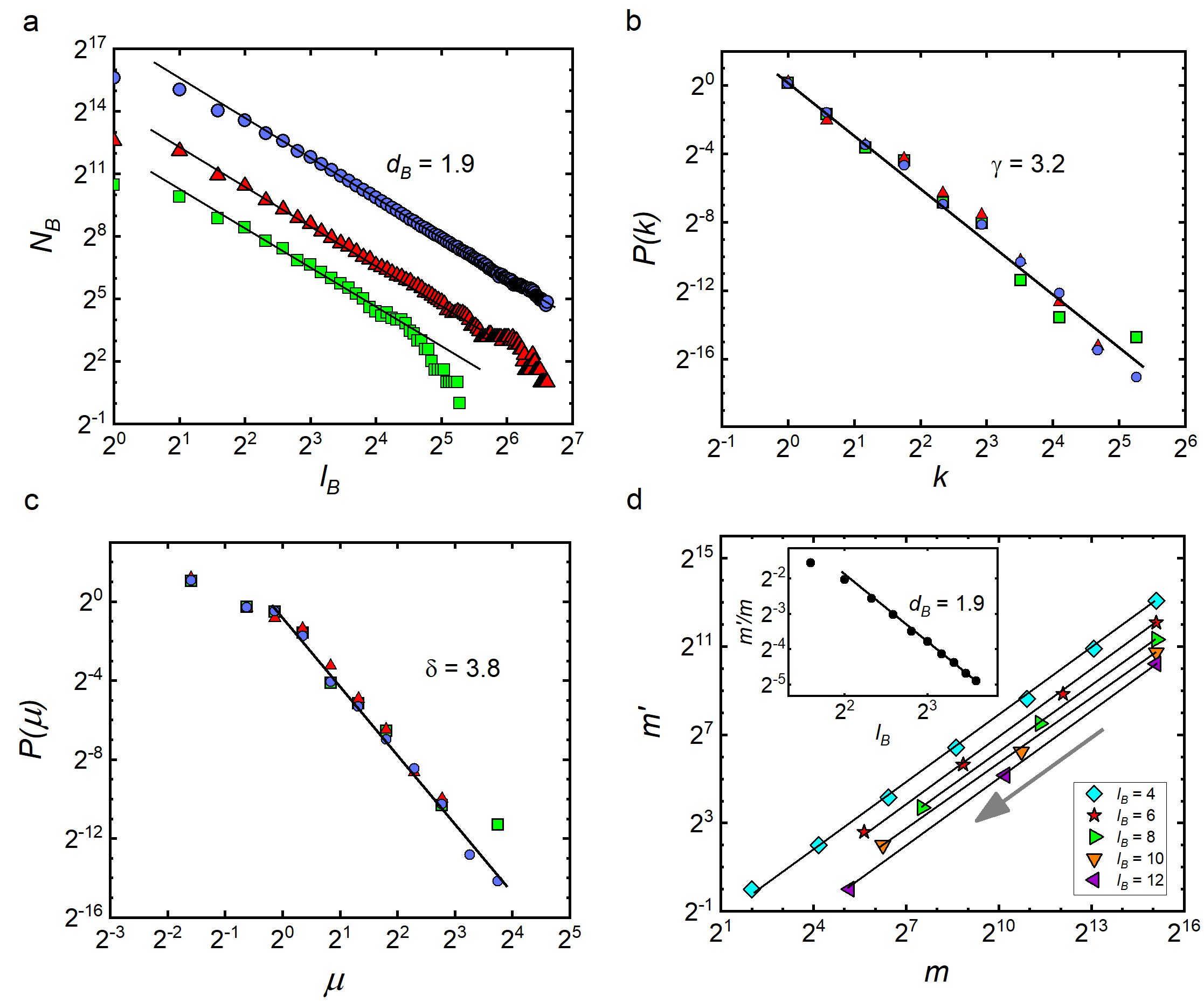}
	\caption{\textbf{Macroscopic characteristics of fractal complex networks}: corresponding to \textbf{(a)} the number of boxes -- $N_B(l_B)$ needed to cover the considered networks as a function of the diameter $l_B$ in the box, \textbf{(b)} the node degree distributions -- $P(k)$, and \textbf{(c)} the distributions of normalized masses of $L$-boxes -- $P(\mu)$, for $L=3$. To construct these graphs, a nested BA network of size $N\simeq 5\cdot 10^4$ and diameter $d=475$ was created (this data are marked with navy circles, see also Tab.~\ref{tabx}). To analyse the self-similarity of the network parts, the original network was covered with boxes of diameter $l_B=40$, and the largest box of size $M\simeq 1.4\cdot 10^3$ was extracted as a new network (this data are marked with green squares). To create the renormalized network of size $N'\simeq 6.1\cdot 10^3$, the original network was covered with boxes of size $l_B=6$, and then each of these $N_B(6)=N'$ boxes was replaced with a supernode (these data are marked with red triangles). In the graph \textbf{(d)}, results of repeated $l_B$-renormalizations of a single box of size $m\simeq 3.5\cdot 10^4$ and diameter $L=300$ are shown, which allow for an alternative determination of the box dimension of the studied fractal network (see the description in the main text of the paper). \label{figmain2}}
\end{figure}

To grasp the relation between Eqs.~(\ref{NB}) and~(\ref{intro3}), it is enough to analyse the meaning of Eq.~(\ref{intro3}), which can be interpreted in two ways. More directly, it states that if one considers \textit{a smaller part of the system}, let's say of size $L'=bL$ (with $b<1$), then $m(L')$, as compared to $m(L)$, is decreased by a factor $\mu(b)=b^{d_f}$, which only depends on $b$. However, this equation also applies to the masses of the system \textit{on two different scales}, or resolutions, which, from a formal point of view, can be treated as two stages of some renormalization procedure applied to that system. (A network-based illustration of these two interpretation schemes is shown in Figs.~\ref{figmain1} and \ref{figmain2}(a-c).) Accordingly, to make Eq.~(\ref{intro3}) more operationalizable, it can be rewritten as:
\begin{equation}\label{intro4}
	m'(L')=b^{d_f}m(L),
\end{equation}
where the notation with the apostrophe is introduced to indicate the relation between the mass of the system before renormalization, $m(L)$, to its mass after renormalization, $m'(L')$. Now, it is easy to see that Eq.~(\ref{NB}) is indeed a special case of Eq.~(\ref{intro4}), with: $d_f=d_B$, $b=l_B^{-1}$, $m(L)=N$, and $m'(L')=N_B(l_B)$, where $L$ and $L'$ stand for diameters of the network before and after renormalization, respectively. 

In what follows, to extend the concept of geometric self-similarity to fractal complex networks, we assume that Eq.~(\ref{intro4}) can be rewritten in the form: 
\begin{equation}\label{main1}
	m'(L',k')=l_B^{-d_B}m(L,k),
\end{equation}
where $d_B$ is the box dimension of fractal networks, whereas $m(L,k)$ and $m'(L',k')$ stand for the number of nodes and supernodes in the same box, before and after its renormalization with boxes of diameter $l_B<L$, respectively. In other words, in Eq.~(\ref{main1}), $m'(L',k')$ is equal to the number of $l_B$-boxes used to cover the initial box of mass $m(L,k)$. As indicated in this equation, during renormalization, when $l_B$-boxes are replaced with supernodes, not only the mass of the initial box changes, but also its diameter (from $L$ to $L'$) and the degree of its hub (from $k$ to $k'$, where $k'$ is the degree of the best-connected $l_B$-box within the initial $L$-box). 

Now, since Eq.~(\ref{main1}), like Eqs.~(\ref{intro3}) and~(\ref{intro4}), defines a generalized homogeneous function \cite{2004bookMcComb} of the form:
\begin{equation}\label{main4}
	m(L,k)=B\,L^\alpha k^\beta,
\end{equation}   
after its substitution into (\ref{main1}), we obtain several scaling relations characterizing fractal networks. The first relation poses: 
\begin{equation}\label{main2}
	L'=L/l_B^{\;d_L}=L/l_B,
\end{equation}
where $d_L=1$ is a direct consequence of the applied renormalization procedure, assuming perfect tiling of the network with boxes of diameter $l_B$ each \cite{2006SongNatPhys}. The second relation has the form of the long-confirmed empirical relation \cite{2005SongNature},
\begin{equation}\label{dk}
	k'=k/l_B^{\;d_k},
\end{equation}   
but in the context of Eq.~(\ref{main1}), which applies to boxes of any diameter $L\geq l_B$, the range of its applicability is much wider than previously thought, which in view of our approach was limited to the case of $L=l_B$ and $L'=1$. Finally, taken together Eqs.~(\ref{main1})-(\ref{dk}) give the following scaling relation:  
\begin{equation}\label{main5}
	d_B=\alpha d_L+\beta d_k=\alpha+\beta d_k,
\end{equation}
which is one of the most important results of this article.

According to Eq.~(\ref{main5}), the box dimension $d_B$ of fractal networks is only determined by the scaling exponents characterizing the microscopic structure of the network at the box level. In particular, as follows from Eq.~(\ref{main1}), beside the method of determining the box dimension, which involves counting non-overlapping boxes (see~Fig.~\ref{figmain2}(a)), $d_B$ can also be obtained by subsequent renormalizations of a single $L$-box with smaller boxes of a given diameter $l_B<L$ (see~Fig.~\ref{figmain2}(d)). The decreasing sequence of renormalized masses of the box obtained in this way: $m,m',m'',\dots,m^{(i)},m^{(i+1)}\dots$, when presented on a graph in the form of points $(m^{(i)},m^{(i+1)})$, can be used to determine the coefficient $l_B^{-d_B}$ in Eq.~(\ref{main1}). Repeating this procedure for different values of $l_B$, a set of points $(l_B,l_B^{-d_B})$ can be obtained which, when fitted with straight line on a double logarithmic scale, gives the same value of $d_B$, as that resulting from the classical method based on Eq.~(\ref{NB}).

The form of Eq.~(\ref{main5}) is also very suggestive. It is the sum of two components, each of which is the product of scaling exponents relating to specific quantities characterizing the mass of the box before and after renormalization. In classical fractals, in which the mass of the box depends only on its linear size $L$, this sum has only one component. For this reason, in classical fractals, the fractal dimension can be determined by one of two methods: the box-covering method or the cluster-growing method, which are equivalent to each other. However, this is not the case of fractal complex network \cite{2005SongNature, 2019WeiPhysA}, where $\alpha$, playing the role of the spreading dimension, only describes how the mass of the box, Eqs.~(\ref{main4}), varies with its diameter:
\begin{equation}\label{main6}
	m(bL,k)=b^{\alpha}m(L,k), 
\end{equation}
where $b>0$. In a similar vein, the second addend in (\ref{main5}), which is further called the mass exponent (in analogy to the degree exponent, $d_k$),
\begin{equation}\label{dm}
	d_m=\beta d_k,
\end{equation}
only characterizes, how the local network \textit{density} (understood as the number of nodes per local area of dimeter $L=L'$) changes as a result of renormalization: 
\begin{equation}\label{main7}
	m'(L,k')=l_B^{-d_m}m(L,k).
\end{equation}

Finally, an observation of great importance for the scaling theory of fractal complex networks (see \hyperref[SecScaleFree]{“Scale-free property”} section) is that when the box mass $m(L,k)$ (\ref{main4}) is divided by the average mass $\langle m\rangle=N/N_B(L)=L^{d_B}$ (\ref{NB}) one gets the normalized mass: 
\begin{equation}\label{mu0}
	\mu(L,k)=\frac{m(L,k)}{\langle m\rangle}=B\,L^{-d_m}k^\beta,
\end{equation}
which turns out to be the invariant of the $l_B$-renormalization procedure, since
\begin{equation}\label{mu1}
	\mu(L,k)=\frac{m(L,k)}{\langle m\rangle}=\frac{l_B^{\,d_B}m'(L',k')}{L^{d_B}}=\frac{m'(L',k')}{L'^{\,d_B}}= \frac{m'(L',k')}{\langle m'\rangle}=\mu'(L',k').
\end{equation}
Indeed, the normalized box mass (\ref{mu0}) is a local network measure that behaves the same regardless of the scale of observation, as the following scaling relations clearly describe:  
\begin{equation}\label{mu3}
	\mu(bL,k)=b^{-d_m}\mu(L,k),
\end{equation} 
and
\begin{equation}\label{mu4}
	\mu'(L,k')=l_B^{-d_m}\mu(L,k).
\end{equation} 
A proper perspective on the meaning of these two relations is gained when comparing them with the corresponding relations for classical fractals, namely Eqs.~(\ref{intro3}) and~(\ref{intro4}). From this perspective, the scaling exponent $d_m$ appears to be the self-similarity dimension of fractal complex networks, which, remarkably, is different from the box dimension $d_B$.

\subsection*{Scale-free property}\label{SecScaleFree}

At this point, we would like to emphasize the lack, in our considerations so far, of scale-free node degree distributions, whose invariance due to the renormalization procedure is considered an attribute of fractal networks \cite{2005SongNature, 2006SongNatPhys, 2009RozenfeldEncyclopedia}. Interestingly, this lack clearly shows the otherwise obvious fact that fractal networks may not have the scale-free property. Nevertheless, when they reveal the property, then both the node degree distribution, $P(k)\sim k^{-\gamma}$, and the box mass distribution, $P(m)\sim m^{-\delta}$, are invariant under the box-covering renormalization procedure, with their invariance being a consequence of the already discussed geometric self-similarity of boxes and the scale-free property of the distribution of normalized masses $P(\mu)\sim\mu^{-\delta}$, from which $P(m)$ inherits its characteristic exponent $\delta$ (see Fig.~\ref{figmain2}(b,c)).

To show this, let us assume that $P(\mu)$ is scale-free:
\begin{equation}\label{new1}
	P(\mu;L)\sim \mu^{-\delta},
\end{equation} 
where, by writing $P(\mu;L)$ instead of $P(\mu)$, we emphasize that all boxes in the network have the same diameter $L$. To clarify, this distribution refers to the normalized masses of non-overlapping boxes of diameter $L$ used to cover the network. Invariance of this distributions in networks after $l_B$-renormalization is due to the property (\ref{mu1}) which implies that $P(\mu;L)=P(\mu';L')$, i.e.
\begin{equation}\label{new5}
	P'(\mu';L')\sim \mu'^{-\delta}.
\end{equation}

Now, having the relationship between $\mu$, $m$ and $k$ (\ref{mu0}) and using it together with Eq.~(\ref{new1}) in the balance equations between the corresponding distributions, i.e. $P(\mu)d\mu=P(m)dm$ and $P(\mu)d\mu=P(k)dk$, it is easy to show that
\begin{equation}\label{Pm}
	P(m;L)\sim m^{-\delta},
\end{equation} 
and
\begin{equation}\label{new3}
	P(k;L)\sim k^{-\gamma},
\end{equation} 
where the characteristic exponent $\gamma$ is given by:
\begin{equation}\label{new4}
	\gamma=1+\beta(\delta-1).
\end{equation} 
The invariance of these distributions in networks after $l_B$-renormalization is obvious due to Eq.~(\ref{new5}).

The above reasoning shows that the geometric self-similarity of the boxes (\ref{mu1})-(\ref{mu4}) and the scale-free distribution of their normalized masses (\ref{new1}) themselves guarantee the invariance of $P(k)$ and $P(m)$ under renormalization. Another consequence of these two assumptions (i.e. self-similarity and scale-freeness), which is not obvious, although it may seem so at first glance, is independence of $P(\mu;L)$~(\ref{new1}) from the diameter of the boxes $L$ (of course, the same applies to $P'(\mu';L')$). In general, this feature can be shown to be true by comparing the numbers of boxes having the same hub nodes when the network is covered with boxes of different diameters. Because the diameter and degree of the hub determine the mass of the box, such a comparison comes down to comparing the number of boxes with the given normalized masses: 
\begin{equation}\label{NB1}
	N_B\!\left(\mu\!(L,k)\right)\,d\!\mu(L,k)=N_B\!\left(\mu\!(bL,k)\right)\,d\!\mu(bL,k),
\end{equation}
where the relationship between the considered masses is determined by Eq.~(\ref{mu3}). Making the appropriate substitutions in this equation, i.e. $N_B(\mu(L,k))=N_B(L)\,P(\mu;L)\sim L^{-d_B}(L^{-d_m}k^\beta)^{-\delta}$ and $N_B(\mu(bL,k))=N_B(bL)\,P(\mu;bL)\sim (bL)^{-d_B}P(\mu;bL)$, cf.~Eqs.~(\ref{NB}),~(\ref{mu0}) and~(\ref{new1}), not only do we confirm that the distribution $P(\mu;bL)$ is scale-free regardless of $b$~(\ref{new1}), but we also obtain a new relation between the scaling exponents:
\begin{equation}\label{delta}
	\delta=1+\frac{d_B}{d_m}.
\end{equation}
Interestingly, using Eqs.~(\ref{dm}) and~(\ref{new4}), the above relation can be easily transformed into the well-known relation \cite{2005SongNature} 
\begin{equation}\label{gamma}
	\gamma=1+\frac{d_B}{d_k}.
\end{equation}

At this point, a natural question to ask is: How is it possible that the relation (\ref{gamma}) was originally derived without having to refer to the self-similarity of the boxes? In fact, as we show below, self-similarity cannot be ignored, and the derivation described in Ref.~\cite{2005SongNature} takes it into account, albeit implicitly. 

More precisely, in the original reasoning leading to Eq.~(\ref{gamma}), one starts with the following equation:
\begin{equation}\label{main8}
	N(k)dk=N'(k')dk',	
\end{equation}
where $N(k)$ (respectively, $N'(k')$) is the number of nodes with $k$ (respectively, $k'$) links in the network before (after) renormalization. Then, substitutions are made in this equation: $N(k)=NP(k)$ and $N'(k')=N'P'(k')$, where $N$ and $N'=N_B(l_B)=Nl_B^{-d_B}$~(\ref{NB}) stand for the number of nodes in the network before and after renormalization, respectively. These substitutions lead to the following density balance equation:
\begin{equation}\label{main9}
	P(k)dk=l_B^{-d_B}P'(k')dk',	
\end{equation}
from which Eq.~(\ref{gamma}) is obtained under the assumptions that both node degree distributions are scale-free with the same scaling exponent, i.e. $P(k)\sim k^{-\gamma}$ and $P'(k')\sim k'^{-\gamma}$, and that Eq.~(\ref{dk}) is met between $k$ and $k'$. It should be emphasized, however, that what underlies validity of these assumptions is the geometric self-similarity of the boxes and the scale-free distribution of their masses. Furthermore, this derivation itself is a special case of more general considerations, in which the starting point is the below equation: 
\begin{equation}\label{NB2}
	N_B\!\left(\mu\!(L,k)\right)\,d\!\mu(L,k)=N_B'\!\left(\mu'\!(L,k')\right)\,d\!\mu'(L,k'),
\end{equation}
whose logic is similar to that behind Eq.~(\ref{NB1}). To explain, the left-hand side of Eq.~(\ref{NB2}) represents the number of boxes with diameter $L$ and hub degree $k$ in the network before renormalization, while the right-hand side is the number of boxes with the same diameter $L$ and hubs of degree $k'$ in the network after $l_B$-renormalization. The numbers of these boxes must match because hubs of degree $k'$ in the network after renormalization arise from those $l_B$-boxes in the network before renormalization that contained hubs of degree $k$. The relation between the masses of the considered boxes is given by Eq.~(\ref{mu4}).

Interestingly, for arbitrary value of $L$, the scaling analysis of Eq.~(\ref{NB2}) leads to the scaling relation~(\ref{delta}). However, when $L\!=\!1$ is assumed, then Eq.~(\ref{NB2}) can transformed to Eq.~(\ref{main8}). In particular, the left hand side of Eq.~(\ref{NB2}) becomes:  $N_B(\mu(1,k))d\mu(1,k)=NP(\mu(1,k))\frac{d\mu(1,k)}{dk}dk=\beta Nk^{-\beta(\delta-1)-1}dk=\beta Nk^{-\gamma}dk=\beta N(k)dk$, where we one by one used Eqs.~(\ref{NB}), (\ref{new1}), (\ref{mu0}), and (\ref{new4}). Similar transformations applied to the right-hand side of Eq.~(\ref{NB2}) result in: $N_B'(\mu'(1,k'))d\mu'(1,k')=\beta N'(k')dk'$, what was to be shown.

\begin{figure}[t]
	\centering
	\includegraphics[width=0.95\columnwidth]{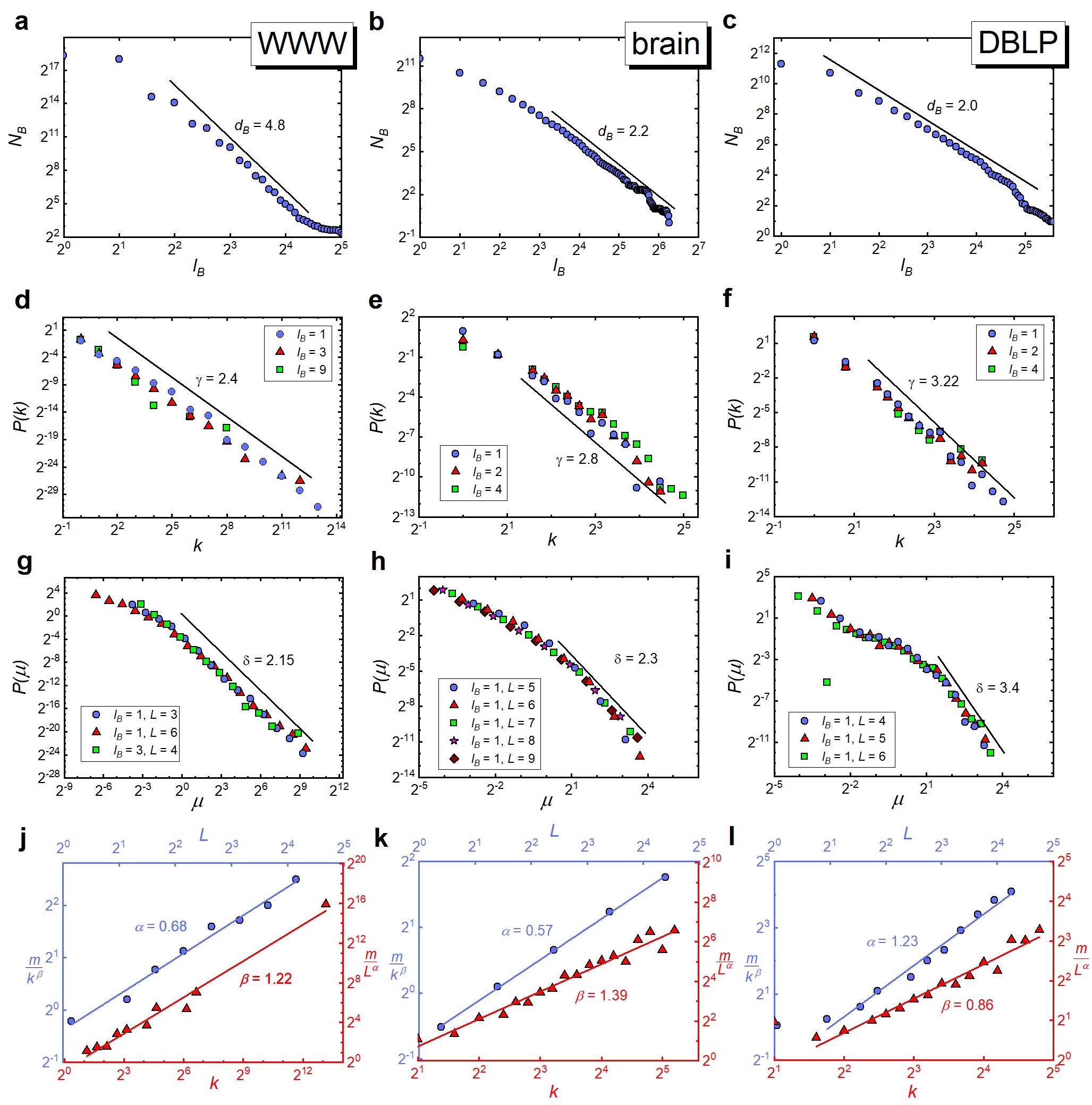}
	\caption{\textbf{Scale-invariant and self-similar scaling in real fractal networks}. The graphs placed in the same column refer to the same network (i.e. WWW, brain and DBLP, respectively, starting from the left), and those placed in the same row to the same scaling relation. In particular, the following graphs show: \textbf{(a-c)} A log-log plot of $N_B$ versus $l_B$ revealing the fractal nature of the studied network according to Eq.~(\ref{NB}). \textbf{(d-f)} Invariance of the node degree distribution $P(k)$ under the renormalization for different box sizes $l_B$ (the case of $l_B=1$ corresponds to the original network). \textbf{(g-i)} Invariance of the normalized mass box distribution $P(\mu)$ (where $L$ represents diameter of the considered boxes). \textbf{(j-l)} Scaling of the masses of boxes according to Eq.~(\ref{main4}). (See the description given in the main text.)\label{figReal}}
\end{figure}

\begin{figure}[t]
	\centering
	\includegraphics[width=0.92\columnwidth]{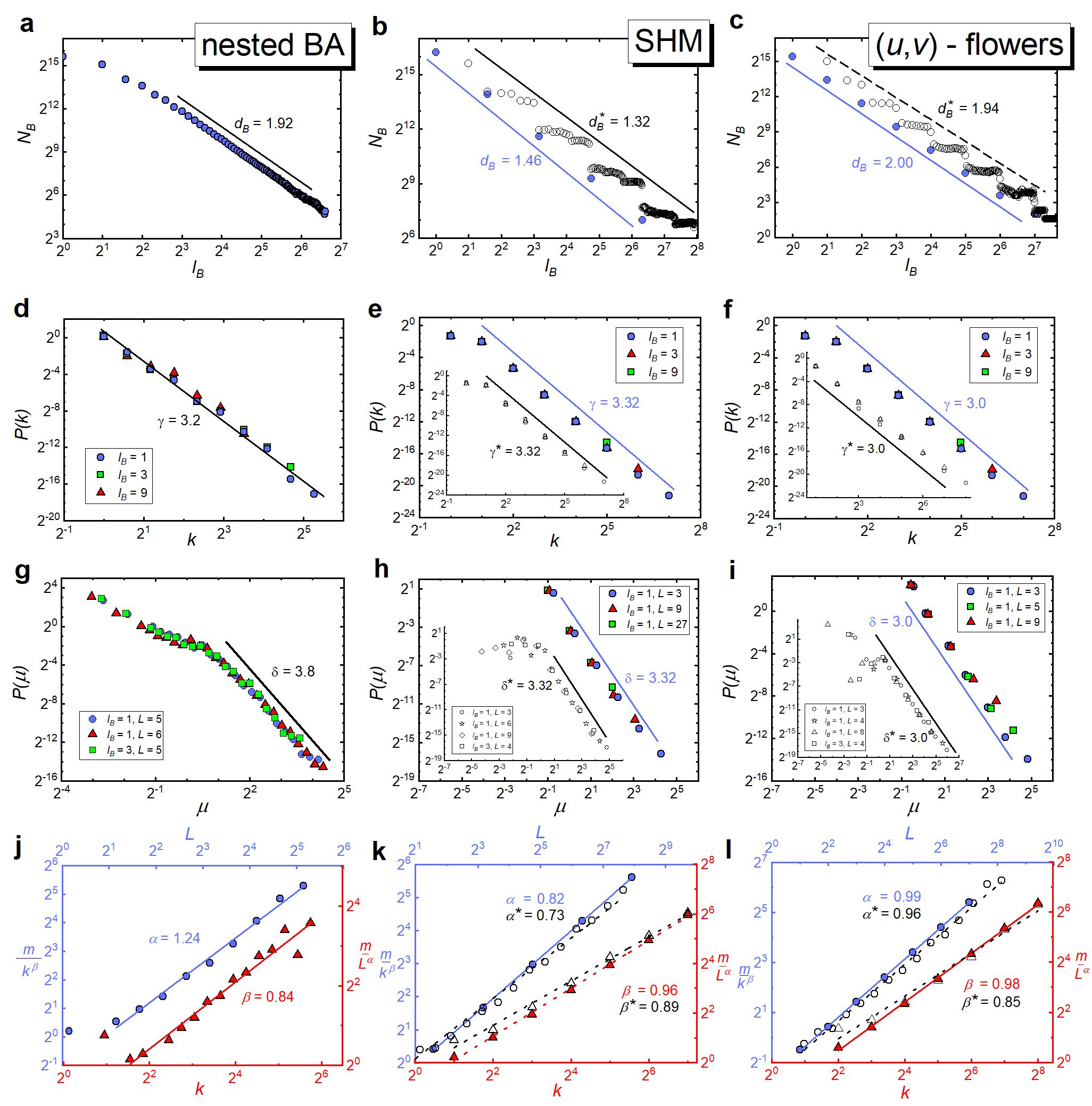}
	\caption{\textbf{Scale-invariant and self-similar scaling in model-based fractal networks}. As in Fig.~\ref{figReal}, the graphs placed in the same column refer to the same model of fractal networks (i.e. the nested BA networks, the SHM model, and (u,v)-flowers, respectively, starting from the left), and those placed in the same row to the same scaling relation. The presentation of data in this figure compared to Fig.~\ref{figReal} differs only in that the graphs relating to deterministic models show two types of points: closed and open. For these models, closed points refer to the box-covering method resulting from their deterministic construction procedure, which uses a significantly smaller number of boxes than the Song's algorithm, whose results correspond to open points (see the main text of the paper for more detailed explanation). Tab.~\ref{tab1} shows the results obtained based on the closed points.}\label{figModel}
\end{figure}

\section*{From microscopic to macroscopic scaling exponents in real and model-based fractal networks}

All scaling exponents discussed in this article, which describe fractal complex networks, can be divided into two groups. The first group refers to the macroscopic characteristics of the network ($d_B$, $\gamma$, and $\delta$), and the second group includes the exponents that characterize the network structure at the microscopic level ($d_k$, $d_m$, $\alpha$ and $\beta$). Interestingly, exponents from both groups are related to each other and, as in the scaling theory of critical phenomena, only a few of them, three to be exact, are independent. The choice of the three fundamental exponents depends on the focus of the study. Here, to validate our results in real and model-based fractal networks, we take the easier to measure macroscopic exponents as independent. This choice results in the following set of test relations, cf.~Eqs. (\ref{delta}) and~(\ref{gamma}): 
\begin{equation}\label{dkdm}
	d_m=\frac{d_B}{\delta-1},\;\;\;\;\;\;\;\;\;\;d_k=\frac{d_B}{\gamma-1},
\end{equation}
and, cf.~Eqs.~(\ref{main5}) and (\ref{dm}): 
\begin{equation}\label{ab}
	\alpha=\frac{\delta-2}{\delta-1}\,d_B,\;\;\;\;\;\;\;\;\;\;\beta=\frac{\gamma-1}{\delta-1},
\end{equation}
of which only the relation for $d_k$ (\ref{dkdm}) has been verified in real \cite{2005SongNature} and model \cite{2006SongNatPhys} networks, and the results of the validation of relations (\ref{ab}) are summarized below.

The real networks analyzed in this paper come from various fields and represent information, social, and biological networks. We analyzed: 1)~a~sample of the WWW with nodes corresponding to web pages and links standing for hyperlinks \cite{1999AlbertNature}; 2)~a~coauthorship network (DBLP), where nodes are scientists and edges are placed between two scientists if they have co-authored a paper \cite{DBLP1,2022FronczakSciRep}; 3)~a~functional brain network, which reflects the correlation between the activity of different areas in the human brain \cite{2012GallosPNAS, 2014ReisNatPhys}. In addition to real networks, we have also analyzed several fractal network models, including our own network model, which is based on nested BA networks \cite{1999BarabasiScience}, the Song-Havlin-Makse (SHM) model \cite{2006SongNatPhys} and (u,v)-flowers \cite{2007RozenfeldNewJPhys}. Detailed information on all these networks (real and synthetic) can be found in \hyperref[SecMethods]{“Methods”} section.

\begin{table*}[!ht]
	\footnotesize
	\caption{\label{tabx}\textbf{Values of the scaling exponents for various fractal networks}. In the table, $N$ is the number of nodes in the analyzed network, $\langle k\rangle$ is the average node degree, and $d$ corresponds to the diameter of the network. The empirical values of the scaling exponents were determined by fitting the appropriate scaling relations to real data and results of numerical simulations, for real and model networks, respectively. The theoretical values of microscopic exponents, which are given in brackets, are of two types. For real networks and for the nested BA model, they were calculated on the basis of appropriate scaling relations using empirical values of macroscopic exponents, and for deterministic models of fractal networks, they were calculated on the basis of derived theoretical relations.}
	\centering
	\label{tab1}
	\begin{tabular}{|c|c|c|c|c|c|c|c|c|}
		\hline
		\textbf{network} & \textbf{$N$} & \textbf{$\langle k\rangle$} & \textbf{$d$} & \textbf{$d_B$} & \textbf{$\gamma$} & \textbf{$\delta$} & \textbf{$\alpha$} & \textbf{$\beta$} \\ \hline\hline
		\textbf{WWW} & 325728 & 4.6 & 46 & 4.8 & 2.4 & 2.2 & 0.68 (0.63) & 1.22 (1.22) \\ \hline
		\textbf{DBLP} & 2523 & 2.5 & 62 & 2.0 & 3.2 & 3.4 & 1.23 (1.17) & 0.86 (0.92)\\ \hline
		\textbf{brain} & 2920 & 4.7 & 77 & 2.2 & 2.8 & 2.3 & 0.57 (0.51) & 1.39 (1.38) \\ \hline
		\textbf{nested BA} & 50000 & 2 & 475 & 1.92 & 3.2 & 3.8 & 1.24 (1.23) & 0.84 (0.79) \\ \hline
		\textbf{SHM model} & 78126 & 2 & 4373 & 1.46 & 3.32 & 3.32 & 0.82 & 0.96 \\
		$s\!=\!2$, $a\!=\!3$, $n\!=\!2s\!+\!1\!=\!5$& ~ & (tree) & ~ & $\left(\frac{\ln n}{\ln a}\simeq 1.46\right)$ & $\left(1+\frac{\ln n}{\ln s}\simeq 3.32\right)$ & $\left(1+\frac{\ln n}{\ln s}\simeq 3.32\right)$ & $\left(\frac{\ln(n/s)}{\ln s}\simeq 0.83\right)$ &  (1) \\ \hline
		\textbf{(u,v)--flowers} & 43692 & 3 & 416 & 2.0 & 3.0 & 3.0 & 0.99 & 0.98 \\ 
		$u\!=\!2$, $v\!=\!2$, $w\!=\!u\!+\!v\!=\!4$ & ~ & ~ & ~ & $\left(\frac{\ln w}{\ln u}=2 \right)$ & $\left(1+\frac{\ln w}{\ln 2}=3 \right)$ & $\left(1+\frac{\ln w}{\ln 2}=3 \right)$ & $\left(\frac{\ln(w/2)}{\ln u}=1 \right)$ & (1) \\ \hline
	\end{tabular}
\end{table*}

Table \ref{tabx} presents the theoretical and empirical values of the scaling exponents of all analyzed networks. The theoretical values, which are given in brackets, are of two types. For the deterministic model-based networks---the SHM model and (u,v)-flowers---their values can be calculated using the appropriate formulas, the details of which are provided in \hyperref[SecMethods]{“Methods”} section (more precisely: \hyperref[MethodsSHM]{“SHM model”} and \hyperref[MethodsFlowers]{“(u,v)-Flowers”} subsections, respectively). For real networks and for the numerical model of nested BA, the theoretical values of $\alpha$ and $\beta$ were calculated from Eqs.~(\ref{ab}) using the empirical values of the macroscopic exponents. 

Correspondingly, the empirical values of the scaling exponents were calculated from Figs.~\ref{figReal} and~\ref{figModel} according to the following protocol (the same for each network): First, we determined the box dimension $d_B$ of these networks resulting from tiling the network with boxes of different sizes $l_B$, see Figs.~\ref{figReal},~\ref{figModel}~(a-c). To this end, we used the algorithm developed by Song et al. \cite{2007SongJStatMech}, and in the case of deterministic models of fractal networks, shown in~Fig.~\ref{figModel}, we additionally analysed the tiling consistent with their deterministic construction procedures, finding that they use a much smaller number of boxes than the Song method. We confirmed that the value of $d_B$ after renormalization (even multiple times) remains the same as before renormalization, see Fig.~\ref{figmain2}(a). We then examined the invariance of distributions $P(k)$ and $P(\mu)$. The given values of $l_B$ refer to the diameter of the boxes that are used to renormalize the network. As already stated, $l_B=1$ refers to the original network - before renormalization. In the case of $P(\mu)$ distributions, the diameters $L$ of the boxes whose mass was studied are also given. With respect to these distributions, the provided values of $l_B$ and $L$ should be read as follows: The relevant distribution $P(\mu)$ refers to the network that was first renormalized with boxes of diameter $l_B$ and then covered with non-overlaying boxes of diameter $L$. Regarding $P(\mu)$, however, due to the low statistical reliability of the data for $l_B,L>1$, in this paper, we only present data for the largest networks (i.e. WWW and model based networks). It should be noted that in all networks we studied, both distributions are scale-invariant, with well-defined characteristic exponents $\gamma$ and $\delta$ (see Figs.~\ref{figReal},~\ref{figModel}~(d-i)). Lastly, having determined the macroscopic scaling exponents: $d_B$, $\gamma$, and $\delta$, we were able to calculate the theoretical values of the local exponents---$\alpha$ and $\beta$, Eqs.~(\ref{ab})---which we used to obtain the adequately rescaled masses of boxes to determine their empirical values (see Figs.~\ref{figReal},~\ref{figModel}~(j-l)). In particular, to obtain the empirical value of $\alpha$, the masses of all the internally connected boxes, obtained during tiling the network with different $l_B$-boxes, were divided by the hub's degree raised to the power of the theoretically obtained $\beta$. Such rescaled masses $m/k^\beta$ were then plotted against the actual diameters of the boxes, $L<l_B$, which had been specified individually for each box. A similar procedure was applied to determine the empirical value of $\beta$. (For more details see the subsection titled  \hyperref[MethodsNumerical]{“Numerical calculation of microscopic scaling exponents”} in the "Methods" section).

Interestingly, in the case of deterministic fractal network models, only the box-covering method which takes into account the network construction procedure while using a smaller number of boxes than Song's method, leads to microscopic exponents consistent with their theoretical predictions (cf. Fig.~\ref{figModel}(k,l) and Tab.~\ref{tabx}). In the case of these networks, the poor performance of Song's method is especially visible in the range of small masses of the $P(\mu)$ distributions (see subset graphs in Fig.~\ref{figModel}(h,i)). We suspect, this latter observation may explain the occurrence of two different scaling behaviours (for small and large $\mu$) in $P(\mu)$ of other fractal networks (cf. Figs.~\ref{figReal}(g-i) and \ref{figModel}(g)).

\section*{Perspectives}

The origins and consequences of fractality are one of the three main research directions in the geometry of complex networks \cite{2021BogunaNatPhys}, next to the hyperbolic geometry of hidden network spaces \cite{2008SerranoPRL, 2010KrioukovPRE} and the geometry induced by dynamic processes in networks \cite{2013BrockmannScience, 2015TaylorNatComm, 2017DeDomenicoPRL}. Although these three geometries, due to the various definitions of distance in each of them, are defined differently, there is no doubt that they must be closely related to each other. While these relationships have yet to be explored, evidence of their existence can be found in our results.

For example, when examining deterministic models of fractal networks (SHM model and (u,v)-flowers, see  \hyperref[MethodsSHM]{“SHM model”} and \hyperref[MethodsFlowers]{“(u,v)-Flowers”} subsections, respectively), we noticed that while macroscopic scaling exponents are very stable in the sense that they do not depend on the box-covering method \cite{2007SongJStatMech, 2021KovacsApplNetwSci}, this may not be the case for microscopic exponents. In particular, in the mentioned models, gathering nodes according to their kinship---which is the most optimal, because it corresponds to the smallest number of boxes---gives the values of microscopic exponents closest to their theoretical predictions. Since the degree of kinship can be thought of as a distance in some metric space---the space of kinship---this observation is important. In fact, the fractality of these models may be considered a feature they inherit from their kinship spaces. Here, natural questions arise, such as whether the fractality of real complex networks may result from the properties of hidden (similarity-based) metric spaces \cite{2012PapadopoulosNature}. Similar studies on community structure confirm the existence of such a relationship \cite{2015ZuevSciRep, 2018GarciaPerezJStatPhys, 2021KovacsSciRep}. The mention of the community structure is not entirely accidental here, because, as the example of the DBLP network shows---in which the removal of weak ties reveals its fractal properties (see also \cite{2013GallosPLoS, 2022FronczakSciRep})---the fat-tailed community size distribution \cite{2005PallaNature, 2010FortunatoPhysRep} may result from the scale-invariant distributions of box masses observed in (not necessarily tree-like) fractal skeletons \cite{2006GohPRL, 2007KimPRE} of these networks.

The second thread that we would like to emphasize concerns the geometry induced by diffusion-like dynamic processes in networks \cite{2013BrockmannScience, 2015TaylorNatComm, 2017DeDomenicoPRL}. In classical fractals, this kind of geometry is closely related to the cluster-growing method of calculating their fractal dimensions, which is actually a way of measuring the distance \cite{1988bookFeder}. In complex networks, establishing an analogous relationship has not been possible so far due to the lack of theoretical foundations distinguishing between the box dimension - $d_B$~(\ref{main1}) (which can be determined by the box covering method) and the spreading dimension - $\alpha$~(\ref{main6}) (which corresponds to the cluster-growing method). It seems that the scaling theory of fractal complex networks presented in this paper has the potential to break this impasse. This is even more likely since in its general findings, with box masses depending not only on the diameter of the boxes but also on the degree of the best-connected node inside the box, the theory refers to the well-established heterogeneous (degree-based) mean-field theory commonly used to study dynamical processes on complex networks \cite{2008bookBarrat}.  

\section*{Data availability}

The datasets used and the complete Python code for all calculations can be obtained from \cite{2023FronczakSM}.

\section*{Methods}\label{SecMethods}

\subsection*{Real and model-based fractal networks analysed in the paper}\label{MethodsData}

The real networks analysed include:

\begin{itemize}
	\item \textbf{WWW network}: The web subset analysed consists of 326 k web pages that are linked if there is a URL link from one page to another \cite{1999AlbertNature}. This dataset has been analysed for fractal properties in many other papers (see e.g. \cite{2005SongNature, 2006SongNatPhys}). It is publicly available in many network repositories (e.g. \cite{WWW}).
	
	\item \textbf{DBLP coauthorship network}: DBLP is a digital library of article records published in computer science \cite{DBLP1, DBLP2}. In this study, we use the 12th version of the dataset (DBLP-Citation-network V12; released in April 2020, which contains information on approximately 4.9~M articles published mostly during the last 20 years). We ourselves processed the raw DBLP data into the form of coauthorship network, from which we extracted the network backbone by imposing a threshold on the minimum number of joint papers ($\geq 25$) two scientists should have. This procedure significantly reduces the size of the studied network (from 2.9~M nodes and 12.5~M links to 2.5~k nodes and 3.2~k edges), but thanks to it the network becomes naturally fractal.
	
	\item \textbf{Human brain networks}: The networks are based on functional magnetic resonance imaging (fMRI). The fMRI data consists of temporal series, known as the blood oxygen level dependent (BOLD) signals, from different brain regions. To build brain networks, the correlations $C_{ij}$ between the BOLD signals are calculated and the two nodes (brain regions) are connected if $C_{ij}$ is greater than some threshold value $T$. In our case we assume $T=0.85$. The brain networks analysed here were used in \cite{2012GallosPNAS, 2014ReisNatPhys} and can be found at \cite{BRAIN}.
\end{itemize}

\begin{figure}
	\centering
	\includegraphics[width=0.5\linewidth]{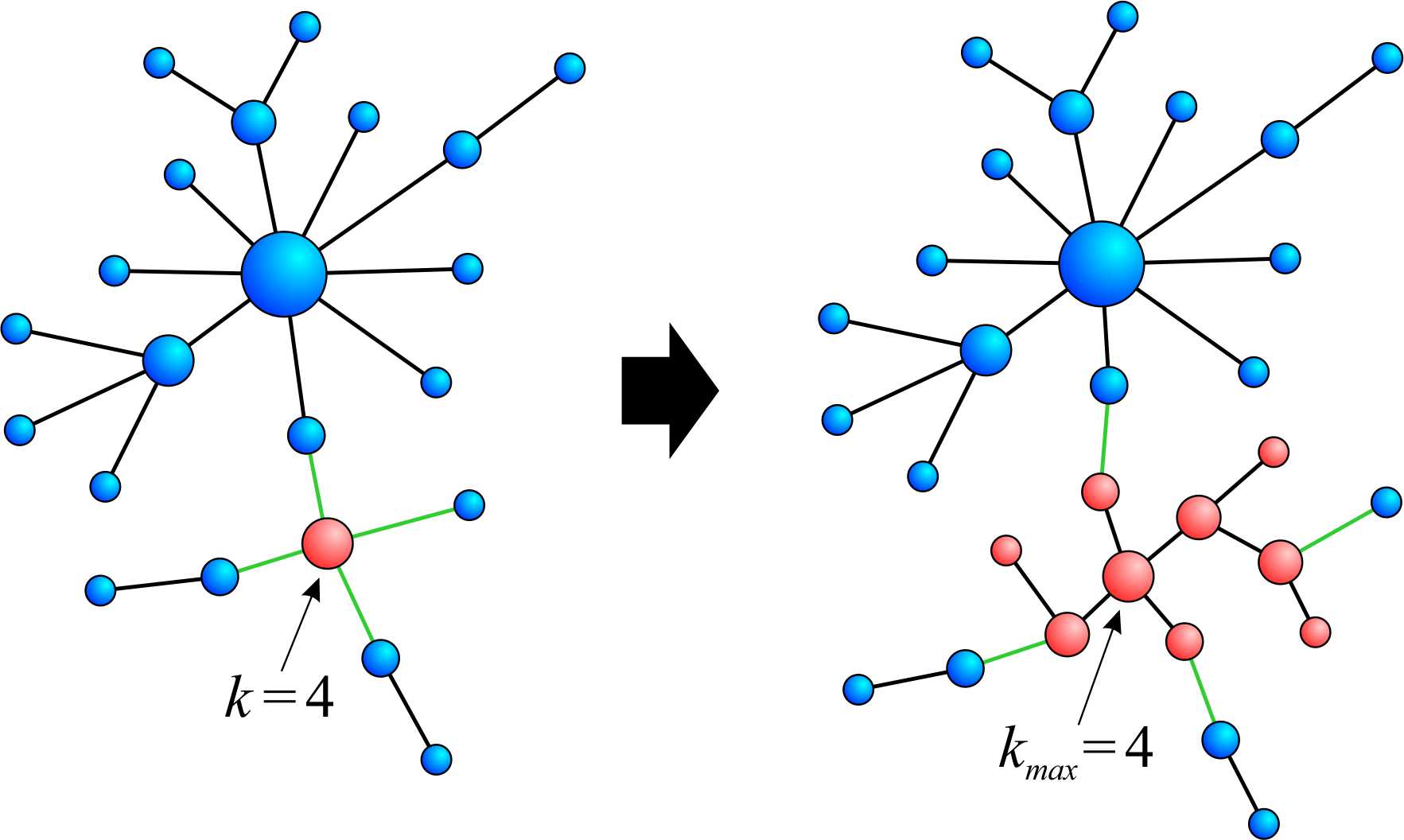}
	\caption{\textbf{Single step of the construction procedure of the nested BA network}. First, one node is chosen with probability that is proportional to its degree, in the figure $k=4$. Then the node is replaced by the corresponding BA network with the best connected node of the same degree $k=4$, as the removed one. Green edges of the removed node are reconnected to randomly selected nodes of the newly created subnetwork.}
	\label{stepBA}
\end{figure}

The studied models of fractal networks include:

\begin{itemize}
	\item \textbf{SHM model}: The details of the model are presented in \hyperref[MethodsSHM]{"SHM model"} subsection, where local scaling exponents for this model were also derived.
	
	\item \textbf{(u,v)-Flowers}: The details of the model are presented in \hyperref[MethodsFlowers]{"(u,v)-Flowers"} subsection, where local scaling exponents for this model were also derived.
	
	\item \textbf{Nested BA networks}: The nested BA network model has three parameters: $N$ - the number of nodes, $k_{max}$ - the degree of the best connected node in the network, and $m$ - the number of edges by which the newly created node connects to the already existing nodes. The network evolution procedure is as follows:
	\begin{enumerate}
		\item First, a BA network with the hub of degree $k_{max}$ is created (that is, the network grows until one of the nodes reaches degree $k_{max}$).
		\item Then, as long as the size of the network is less than $N$ (see Fig.~\ref{stepBA}):
		\begin{enumerate}
			\item a node is chosen proportionally to its degree $k$ and it is replaced with a BA subnetwork with the largest node degree $k$;
			\item edges that were connected to the removed node are reconnected to randomly selected nodes of the newly created subnetwork.
		\end{enumerate}
	\end{enumerate}
\end{itemize}

\subsection*{Numerical calculation of microscopic scaling exponents}\label{MethodsNumerical}

\begin{figure*}[t]
	\centering
	\includegraphics[width=0.9\columnwidth]{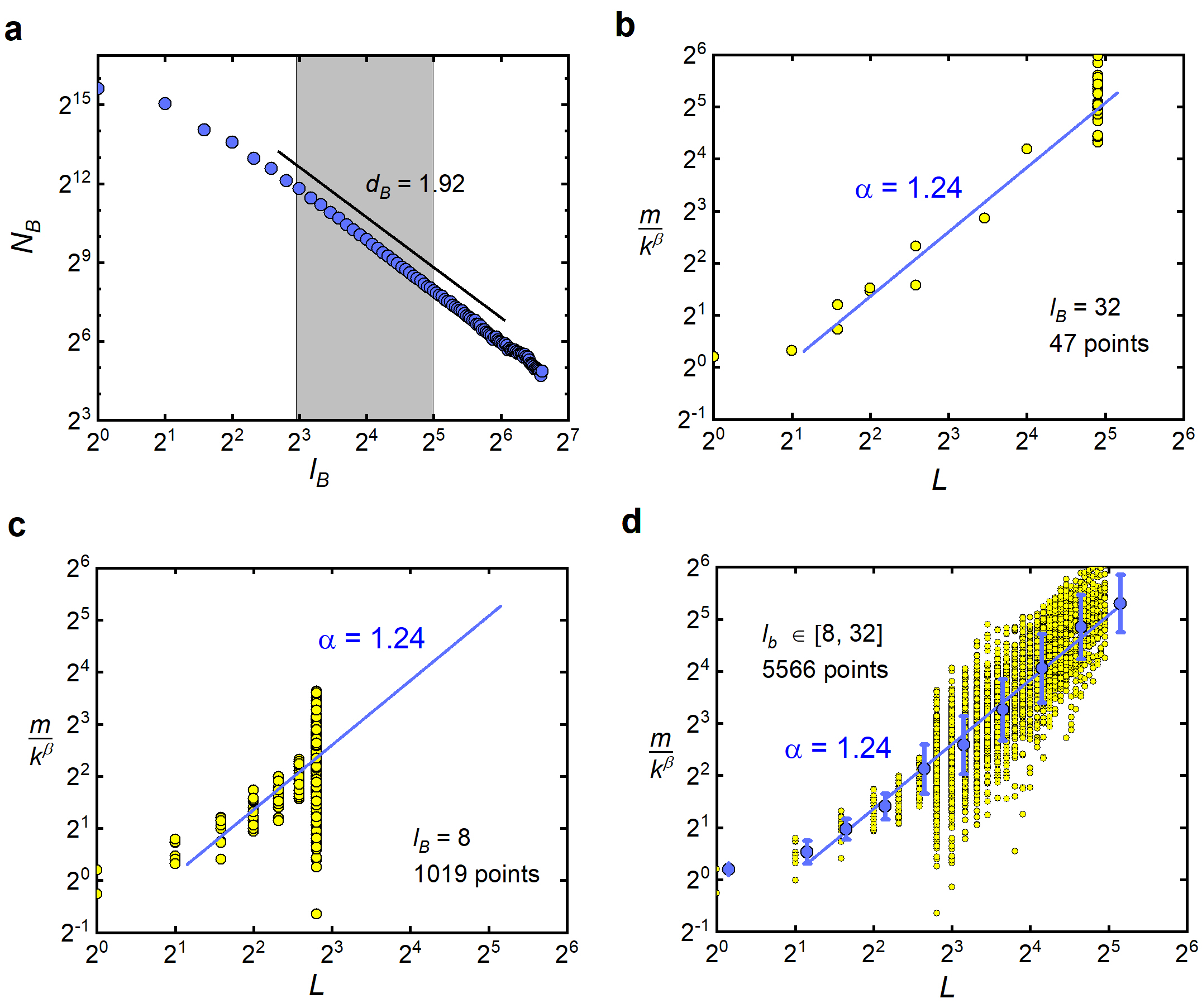}
	\caption{\textbf{Numerical calculation of microscopic scaling exponents.} Detailed description of this figure is given in the text.}\label{figmet}
\end{figure*}

In Figs \ref{figReal}(j-l) and \ref{figModel}(j-l)  we presented the microscopic scaling exponents $\alpha$ and $\beta$. Their values result from fitting a straight line to the set of points marked with blue circles and red triangles, respectively. These points represent the geometric mean of logarithmically equal sized bins of the original data which were obtained as follows. First, we estimated the range of $l_B$ for which dependence of $N_B$ on $l_B$ is linear in log-log scale. In Figure \ref{figmet}(a), which shows an example analysis of nested BA model, this range is indicated by a gray rectangle. Then, for each $l_B$ in this range, we performed box covering, obtaining a triple of values $(m,L,k)$ for each box. Based on the set of such triples and the theoretically calculated value of the $\alpha$ or $\beta$ exponent, a set of points $(k, m/L^\alpha)$ or $(L,m/k^\beta)$ was created, respectively. In Fig. \ref{figmet}(d), the later set is represented by yellow circles. Having this raw data, logarithmic binning has been performed and geometric mean has been calculated for each bin. Fig. \ref{figmet}(d), in addition to the blue points denoting the geometric mean, also shows the geometric standard deviation and the fitted line, whose slope corresponds to the $\alpha$ value we are looking.

If we restrict our analysis to one specific value of $l_B$ (for example, in Fig. \ref{figmet}(b) we take $l_B = 32$, while in Fig. \ref{figmet}(c) we take $l_B = 8$), we end up with a much smaller set of triples that would not allow for a reliable results. These limitations are particularly severe for small real, DBLP, and brain networks, with sizes $N<3\cdot 10^3$ each. The blue lines in Fig. \ref{figmet}(b) and \ref{figmet}(c) are not a result of fitting but are shown for comparison purposes only.

\subsection*{Microscopic scaling exponents for deterministic fractal network models}\label{MethodsModels}

In this section, we derive exact formulas for microscopic scaling exponents ($\alpha$ and $\beta$) characterising deterministic fractal network model.

\subsubsection*{SHM model}\label{MethodsSHM}

In the Song-Havlin-Makse (SHM) model \cite{2006SongNatPhys}, at $t=0$, the network starts to grow from two nodes connected by one link. Then, during subsequent, $t+1$, time steps, next $(t+1)$-generations of the network recursively emerge, in which: $s$ new nodes are attached to the endpoints of each link of the previous $t$-generation, old links are removed from the network, and new links are created in place of those removed, which connect pairs of offspring-nodes attached to the endpoints of the deleted ones. 

As a result of this construction procedure, in successive generations, every node $i$ increases its degree multiplicatively: 
\begin{equation}\label{SHM*ki}
	k_i(t, t_i)=s\,k_i(t-1, t_i)=s^{\Delta t},
\end{equation}
where 
\begin{equation}\label{SHM*Dt}
	\Delta t=t-t_i,
\end{equation}
is the time that has elapsed since the node appeared in the network for the first time, at time $t_i$, assuming that its initial degree \footnote{In fact, the initial degree of one newcomer out of $s$ is $k_i(t_i,t_i)=2$. However, since the nodes' initial degrees does not affect our further calculations, we will not be concerned with this minor oversight.} was equal to $k_i(t_t,t_i)=1$.

It is also easy to see that a similar multiplicative dynamics is also shown by the diameter $L_i$ and mass $M_i$ of the \textit{largest} box where the $i$-th node of degree $k_i(t,t_i)$ acts as a hub \footnote{Such a box consists of all nodes that can be treated as the offspring of $i$, for which the $i$-th node is the parent, grandparent, great-grandfather, etc.}. Specifically, the maximum diameter of such a box is given by:   
\begin{equation}\label{SHM*Li}
	L_i(t,t_i)=a\,L_i(t-1,t_i)=a^{\Delta t}, 
\end{equation}
with \footnote{The value of $a$ results from the construction procedure of the model, since removing the old edges and replacing them with new ones, which are created only between newly added nodes, is formally equivalent to replacing the old edges with paths of length 3 \cite{2006SongNatPhys}.} $a=3$ and $L_i(t_i,t_i)=1$, whereas its mass satisfies the following recurrence relation:
\begin{equation}\label{SHM*Mi}
	M_i(t,t_i)=nM_i(t-1,t_i)=n^{\Delta t},
\end{equation}
where \footnote{The reasoning behind the factor $n$ is the following: at time $t$, the largest box that can be created around the node $i$ consists of the same nodes that formed its largest box at time $t-1$ (the number of which is: $M_i(t-1,t_i)$) and all descendants of those nodes that were created in the last time step (the number of which, due to the tree-like structure of the network, is: $2s(M_i(t-1,t_i)-1)\simeq 2sM_i(t-1,t_i)$). This leads to the following relation: $M_i(t,t_i)=M_i(t-1,t_i)+2sM_i(t-1,t_i)$, which is equivalent to the first part of Eq.~(\ref{SHM*Mi}).} $n=2s+1$ and $M_i(t_i,t_i)=1$. 

At this point, it is worth noting a few remarks regarding Eqs.~(\ref{SHM*Li}) and~(\ref{SHM*Mi}). 

First, Eq.~(\ref{SHM*Li}), when applied to the largest boxes with the oldest nodes, i.e. those from which the network's evolution began, shows how the diameter of the entire network changes in subsequent generations: 
\begin{eqnarray}\label{SHM*Lt}
	L(t)=L_i(t,0)=a\,L_i(t-1,0)=a\,L(t-1). 
\end{eqnarray}
Correspondingly, by applying Eq.~(\ref{SHM*Mi}) to these boxes, one finds the analogous formula for the total number of nodes in the network: 
\begin{eqnarray}\label{SHM*Nt}
	N(t)=M_i(t,0)=n\,M_i(t-1,0)=n\,N(t-1).
\end{eqnarray}
In Ref.~\cite{2006SongNatPhys}, the above recurrence relations, Eqs.~(\ref{SHM*Lt}) and~(\ref{SHM*Nt}), together with Eq.~(\ref{SHM*ki}), were used to derive exact expressions for the box dimension of the considered fractal network model, cf.~Eq.~(1):
\begin{equation}\label{SHM*dB}
	d_B=\frac{\ln n}{\ln a},
\end{equation}
and for its degree exponent, cf.~Eq.~(8): 
\begin{equation}\label{SHM*dk}
	d_k=\frac{\ln s}{\ln a},
\end{equation}
thus enabling verification of the scaling relation (17), according to which, in this model, the characteristic exponent of the degree distribution is given by:
\begin{equation}\label{SM*dk}
	\gamma=1+\frac{d_B}{d_k}=1+\frac{\ln n}{\ln s}.
\end{equation}

The second remark regarding Eqs.~(\ref{SHM*Li}) and~(\ref{SHM*Mi}) it that boxes containing hubs of degrees $k_i(t,t_i)>1$ may have diameters and masses smaller than $L_i(t,t_i)$ and $M_i(t,t_i)$, respectively. For example, when the diameter of the $i$-th box is $l_i=1$, then the box is confined to the node itself and as a result its mass is equal to $m_i=1$. Similarly, when $l_i=a=3$, then the box, apart from the hub itself, also contains all its neighbours, making the mass of the box equal to $m_i=1+k_i\simeq k_i$. More generally, the diameter of the $i$-th box can be equal to:
\begin{equation}\label{SHM*li}
	l_i=a^\tau,
\end{equation}
where
\begin{equation}
	0\leq\tau\leq\Delta t,
\end{equation}
with the value of $\tau$ affecting its mass, which can be determined from \footnote{In fact, the rationale behind Eq.~(\ref{SHM*mi1}) is the same as for Eq.~(\ref{SHM*Mi}). The only difference between $M_i$ and $m_i$ is that the initial condition for the multiplicative growth of the latter is $m_i=1+k_i(t-\tau,t_i)\simeq k_i(t-\tau,t_i)$ and not just $M_i(t_i,t_i)=1$.}:
\begin{equation}\label{SHM*mi1}
	m_i=n^\tau k_i(t-\tau,t_i).
\end{equation}

Now, substituting Eq.~(\ref{SHM*ki}) into~(\ref{SHM*mi1}), one gets:
\begin{eqnarray}\label{SHM*mi2}
	m_i&=&\left(\frac{n}{s}\right)^\tau k_i(t,t_i).
\end{eqnarray}
Then, using Eq.~(\ref{SHM*li}) in~(\ref{SHM*mi2}), one obtains the following relation for the mass of the box as a function of its diameter and hub's degree, cf.~Eq.~(9):
\begin{equation}\label{SHM*mi3}
	m_i=l_i^{\alpha}k_i^{\beta},
\end{equation}
where the local scaling exponents are given by:
\begin{eqnarray}\label{SHM*alpha}
	\alpha=\frac{\ln n-\ln s}{\ln a},
\end{eqnarray}
and
\begin{eqnarray}\label{SHM*beta}
	\beta=1.
\end{eqnarray}	
It is easy to see that, together with the previously obtained expressions for $d_B$ (\ref{SHM*dB}) and $d_k$ (\ref{SHM*dk}), the obtained above expressions for $\alpha$ and $\beta$ satisfy the scaling relation: $d_B=\alpha+d_k\,\beta$, Eq.~(10), which has been derived in the main text of the paper.

\begin{figure*}[t]
	\centering
	\includegraphics[width=0.9\columnwidth]{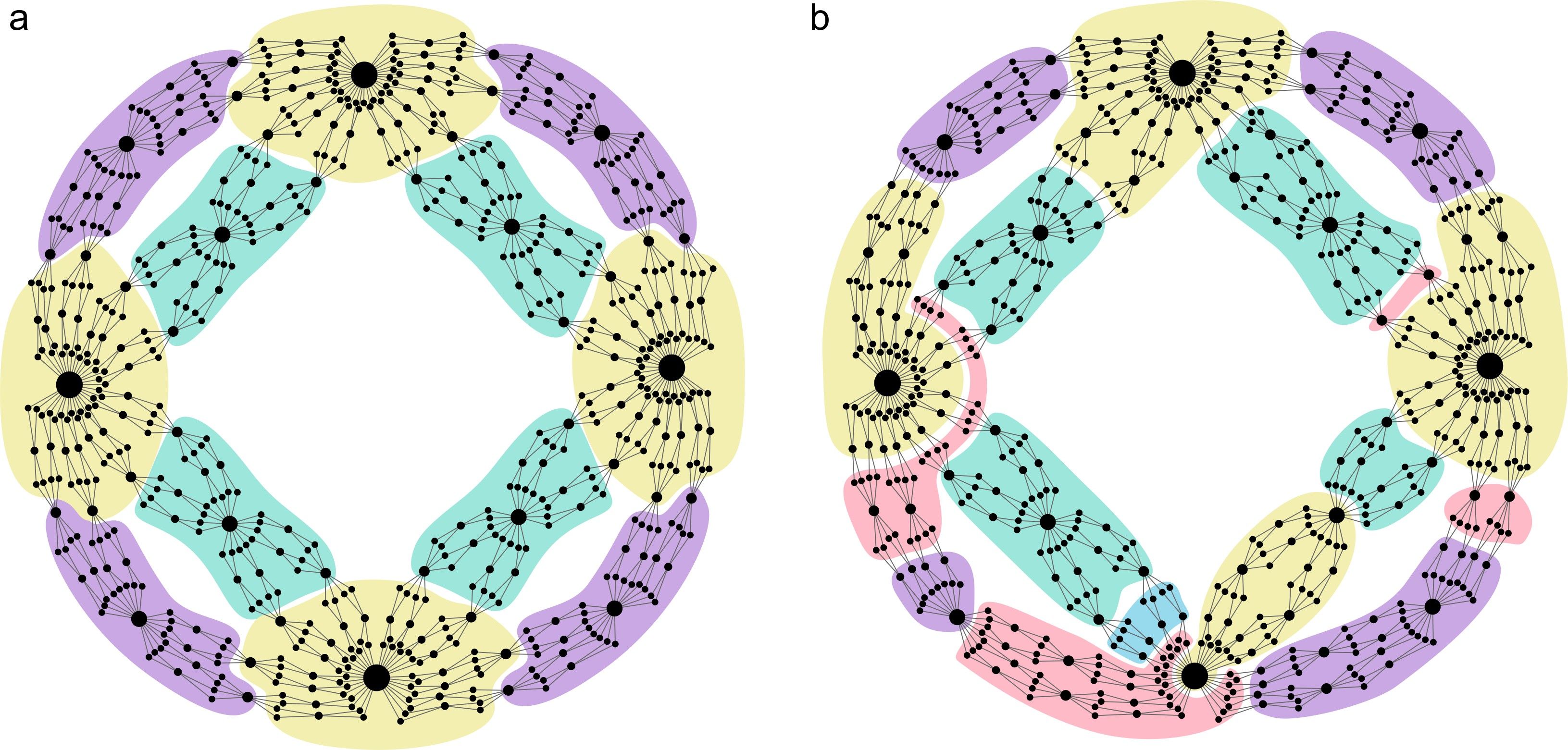}
	\caption{\textbf{(2,2)-Flowers of generation $t=5$}. \textbf{(a)} The network was covered with boxes of diameter $l_B=8$ according to the degree of kinship of the nodes. \textbf{(b)} The node is covered with boxes of diameter $l_B=8$ according to the algorithm developed by Song et al.~\cite{2007SongJStatMech}. It is clear that in the studied deterministic network, Song's random algorithm performs worse than the covering according to the kinship space.\label{flowersboxes}}
\end{figure*}

\subsubsection*{(u,v)-Flowers}\label{MethodsFlowers}

In the deterministic fractal network model called (u,v)-flowers \cite{2007RozenfeldNewJPhys}, networks start to grow, at $t=0$, from two nodes, so-called \textit{initial hubs}, connected by one link. Then, subsequent $(t+1)$-generations of the model are obtained from $t$-generations by replacing each link by two parallel paths of $u>1$ and $v\geq u$ links long. An essential and not obvious at first glance property of this construction procedure is its equivalence to another procedure in which to obtain $(t+1)$-generation one produces $w=u+v$ copies of the previous $t$-generation and then joins the copies at their initial hubs.   

From the second method of constriction, it is easy to see \cite{2007RozenfeldNewJPhys} that the number of links in (u,v)-flowers of generation $t>0$ is given by:
\begin{equation}\label{uvEt}
	E_t=wE_{t-1}=w^t,
\end{equation}
the number of nodes is:
\begin{equation}\label{uvNt}
	N_t=wN_{t-1}-w\;\propto\;w^t,
\end{equation}
and the diameter of the networks grows as:
\begin{equation}\label{uvLt}
	L_t\propto u^t.
\end{equation}
Furthermore, by construction, the networks have only nodes of degree
\begin{equation}\label{uvk}
	k_i=2^n, 
\end{equation}
where $n=1,2,\dots,t$, and their node degree distribution is scale-free (2) with the characteristic exponent equal to:
\begin{equation}\label{uvgamma}
	\gamma=1+\frac{\ln w}{\ln 2}.
\end{equation}
It was also shown that the box dimension (1) of (u,v)-flowers is: 
\begin{equation}\label{uvdB}
	d_B=\frac{\ln w}{\ln u},
\end{equation}
and their degree exponent (8) is:
\begin{equation}\label{uvdk}
	d_k=\frac{\ln 2}{\ln u},
\end{equation}
in accordance with the scaling relation (14). 

In what follows, we show that the local scaling exponents, $\alpha$ and $\beta$ (9), of the model are given by:
\begin{equation}\label{uvalpha}
	\alpha=\frac{\ln w-\ln 2}{\ln u},
\end{equation}
and
\begin{equation}\label{uvbeta}
	\beta=1,
\end{equation}
respectively, so their values satisfy the scaling relation (10).

We first consider the scaling exponent $\beta$. From Eq.~(9), it follows that if the degree of the hub inside the box increases $x$ times, then the mass of the box will increase $x^\beta$ times. Correspondingly, the second method of construction of (u,v)-flowers assumes that in successive generations of these networks, the degrees of the initial hubs double, i.e. $x=2$, which is due to the merger of two initial hubs from two copies of the network of the previous generation. Moreover, since the merged copies are identical, the masses of the boxes with the initial hubs also double, i.e. $x^\beta=2$. Thus, we come to the conclusion that the masses of the boxes are proportional to the degrees of their hubs, which gives $\beta=1$, i.e. Eq.~(\ref{uvbeta}).

To find $\alpha$, we again consider boxes with the initial hubs of degree $k_i=2^t$, Eq.~(\ref{uvk}), in networks of generation $t>0$. Such boxes can be of various diameters. For example, when the diameter of the box is twice the diameter $L_{(t-1)}$ of the network of $(t-1)$-generation, then the mass of the box is twice the number of nodes $N_{(t-1)}$ in the network of $(t-1)$-generation. In general, when the box has a diameter of $2L_n$ (with $0<n<t$), then its mass is equal to (cf. Fig.~\ref{flowersboxes}a):
\begin{equation}\label{uvm1}
	m_i(2L_n,2^t)=2^{t-n}N_n\propto 2^{t-n}w^n.
\end{equation}
Comparing the above relationship with Eq.~(9) one gets:
\begin{equation}\label{uvm2}
	m_i(2L_n,2^t)\propto 2^t\left(\frac{w}{2}\right)^n=2^t(u^t)^\alpha,
\end{equation}
where $\alpha=(\ln w-\ln 2)/\ln u$, cf. Eq.~(\ref{uvalpha}).

\bibliography{fractal1}

\begin{thebibliography}{10}
\expandafter\ifx\csname url\endcsname\relax
  \def\url#1{\texttt{#1}}\fi
\expandafter\ifx\csname urlprefix\endcsname\relax\def\urlprefix{URL }\fi
\providecommand{\bibinfo}[2]{#2}
\providecommand{\eprint}[2][]{\url{#2}}

\bibitem{2005SongNature}
\bibinfo{author}{Song, C.}, \bibinfo{author}{Havlin, S.} \&
  \bibinfo{author}{Makse, H.~A.}
\newblock \bibinfo{title}{Self-similarity of complex networks}.
\newblock \emph{\bibinfo{journal}{Nature}} \textbf{\bibinfo{volume}{433}},
  \bibinfo{pages}{392--395} (\bibinfo{year}{2005}).

\bibitem{2006SongNatPhys}
\bibinfo{author}{Song, C.}, \bibinfo{author}{Havlin, S.} \&
  \bibinfo{author}{Makse, H.~A.}
\newblock \bibinfo{title}{Origins of fractality in the growth of complex
  networks}.
\newblock \emph{\bibinfo{journal}{Nat. Phys.}} \textbf{\bibinfo{volume}{2}},
  \bibinfo{pages}{275--281} (\bibinfo{year}{2006}).

\bibitem{2009RozenfeldEncyclopedia}
\bibinfo{author}{Rozenfeld, H.~D.}, \bibinfo{author}{Gallos, L.~K.},
  \bibinfo{author}{Song, C.},  \& \bibinfo{author}{Makse, H.~A.}
\newblock \emph{\bibinfo{title}{Fractal and transfractal scale-free networks}}
  (\bibinfo{publisher}{Springer}, \bibinfo{address}{New York},
  \bibinfo{year}{2009}).

\bibitem{2020bookRosenberg}
\bibinfo{author}{Rosenberg, E.}
\newblock \emph{\bibinfo{title}{Fractal Dimensions of Networks}}
  (\bibinfo{publisher}{Springer}, \bibinfo{year}{2020}).

\bibitem{2021WenFusion}
\bibinfo{author}{Wen, T.} \& \bibinfo{author}{Cheong, K.~H.}
\newblock \bibinfo{title}{The fractal dimension of complex networks: a review}.
\newblock \emph{\bibinfo{journal}{Inf. Fusion}} \textbf{\bibinfo{volume}{73}},
  \bibinfo{pages}{87--102} (\bibinfo{year}{2021}).

\bibitem{2008RadicchiPRL}
\bibinfo{author}{Radicchi, F.}, \bibinfo{author}{Ramasco, J.~J.},
  \bibinfo{author}{Barrat, A.} \& \bibinfo{author}{Fortunato, S.}
\newblock \bibinfo{title}{Complex networks renormalization: flows and fixed
  points}.
\newblock \emph{\bibinfo{journal}{Phys. Rev. Lett.}}
  \textbf{\bibinfo{volume}{101}}, \bibinfo{pages}{148701}
  (\bibinfo{year}{2008}).

\bibitem{2010RozenfeldPRL}
\bibinfo{author}{Rozenfeld, H.}, \bibinfo{author}{Song, C.} \&
  \bibinfo{author}{Makse, H.~A.}
\newblock \bibinfo{title}{Small-world to fractal transition in complex
  networks: a renormalization group approach}.
\newblock \emph{\bibinfo{journal}{Phys. Rev. Lett.}}
  \textbf{\bibinfo{volume}{104}}, \bibinfo{pages}{025701}
  (\bibinfo{year}{2010}).

\bibitem{2007KimNewJPhys}
\bibinfo{author}{Kim, J.~S.}, \bibinfo{author}{Goh, K.-I.},
  \bibinfo{author}{B., K.} \& \bibinfo{author}{Kim, D.}
\newblock \bibinfo{title}{Fractality and self-similarity in scale-free
  networks}.
\newblock \emph{\bibinfo{journal}{New J. Phys.}} \textbf{\bibinfo{volume}{9}},
  \bibinfo{pages}{177} (\bibinfo{year}{2007}).

\bibitem{2008KimJKorean}
\bibinfo{author}{Kim, J.~S.}, \bibinfo{author}{B., K.}, \bibinfo{author}{Kim,
  D.} \& \bibinfo{author}{Goh, K.-I.}
\newblock \bibinfo{title}{Self-similarity in fractal and non-fractal networks}.
\newblock \emph{\bibinfo{journal}{J. Korean Phys. Soc.}}
  \textbf{\bibinfo{volume}{52}}, \bibinfo{pages}{350} (\bibinfo{year}{2008}).

\bibitem{1997bookDubrulle}
\bibinfo{author}{Dubrulle, B.}, \bibinfo{author}{Graner, F.} \&
  \bibinfo{author}{Sornette, D.}
\newblock \emph{\bibinfo{title}{Scale Invariance and Beyond}}
  (\bibinfo{publisher}{Springer Berlin}, \bibinfo{address}{Heidelberg},
  \bibinfo{year}{1997}).

\bibitem{2006bookSornette}
\bibinfo{author}{Sornette, D.}
\newblock \emph{\bibinfo{title}{Critical Phenomena in Natural Sciences}}
  (\bibinfo{publisher}{Springer Berlin}, \bibinfo{address}{Heidelberg},
  \bibinfo{year}{2006}), \bibinfo{edition}{2} edn.

\bibitem{2005YookPRE}
\bibinfo{author}{Yook, S.-H.}, \bibinfo{author}{Radicchi, F.} \&
  \bibinfo{author}{Meyer-Ortmanns, H.}
\newblock \bibinfo{title}{Self-similar scale-free networks and
  disassortativity}.
\newblock \emph{\bibinfo{journal}{Phys. Rev. E}} \textbf{\bibinfo{volume}{72}},
  \bibinfo{pages}{045105} (\bibinfo{year}{2005}).

\bibitem{2006GohPRL}
\bibinfo{author}{Goh, K.-I.}, \bibinfo{author}{Salvi, G.}, \bibinfo{author}{B.,
  K.} \& \bibinfo{author}{Kim, D.}
\newblock \bibinfo{title}{Skeleton and fractal scaling in complex networks}.
\newblock \emph{\bibinfo{journal}{Phys. Rev. Lett.}}
  \textbf{\bibinfo{volume}{96}}, \bibinfo{pages}{018701}
  (\bibinfo{year}{2006}).

\bibitem{2007KimPRE}
\bibinfo{author}{Kim, J.~S.} \emph{et~al.}
\newblock \bibinfo{title}{Fractality in complex networks: Critical and
  supercritical skeletons}.
\newblock \emph{\bibinfo{journal}{Phys. Rev. E}} \textbf{\bibinfo{volume}{75}},
  \bibinfo{pages}{016110} (\bibinfo{year}{2007}).

\bibitem{2007KitsakPRE}
\bibinfo{author}{Kitsak, M.} \emph{et~al.}
\newblock \bibinfo{title}{Betweenness centrality of fractal and nonfractal
  scale-free model netwroks and tests on real networks}.
\newblock \emph{\bibinfo{journal}{Phys. Rev. E}} \textbf{\bibinfo{volume}{75}},
  \bibinfo{pages}{056115} (\bibinfo{year}{2007}).

\bibitem{2008GallosPRL}
\bibinfo{author}{Gallos, L.~K.}, \bibinfo{author}{Song, C.} \&
  \bibinfo{author}{Makse, H.~A.}
\newblock \bibinfo{title}{Scaling of degree correlations in scale-free
  networks}.
\newblock \emph{\bibinfo{journal}{Phys. Rev. Lett.}}
  \textbf{\bibinfo{volume}{100}}, \bibinfo{pages}{248701}
  (\bibinfo{year}{2008}).

\bibitem{2016WeiPRE}
\bibinfo{author}{Wei, Z.-W.} \& \bibinfo{author}{Wang, B.-H.}
\newblock \bibinfo{title}{Emergence of fractal scaling in complex networks}.
\newblock \emph{\bibinfo{journal}{Phys. Rev. E}} \textbf{\bibinfo{volume}{94}},
  \bibinfo{pages}{032309} (\bibinfo{year}{2016}).

\bibitem{2017FujikiEPJB}
\bibinfo{author}{Fujiki, Y.}, \bibinfo{author}{Mizutaka, S.} \&
  \bibinfo{author}{Yakubo, K.}
\newblock \bibinfo{title}{Fractality and degree correlations in scale-free
  networks}.
\newblock \emph{\bibinfo{journal}{Eur. Phys. J. B}}
  \textbf{\bibinfo{volume}{90}}, \bibinfo{pages}{1--9} (\bibinfo{year}{2017}).

\bibitem{2023ZakarNetSci}
\bibinfo{author}{Zakar-Poly\'{a}k, E.}, \bibinfo{author}{Nagy, M.} \&
  \bibinfo{author}{Molontay, R.}
\newblock \bibinfo{title}{Towards a better understanding of the characteristics
  of fractal networks}.
\newblock \emph{\bibinfo{journal}{Appl. Netw. Sci.}}
  \textbf{\bibinfo{volume}{8}}, \bibinfo{pages}{17} (\bibinfo{year}{2023}).

\bibitem{2007RozenfeldNewJPhys}
\bibinfo{author}{Rozenfelf, H.~D.}, \bibinfo{author}{Havlin, S.} \&
  \bibinfo{author}{ben Avraham, D.}
\newblock \bibinfo{title}{Fractal and transfractal recursive scale-free nets}.
\newblock \emph{\bibinfo{journal}{New J. Phys.}} \textbf{\bibinfo{volume}{9}},
  \bibinfo{pages}{175} (\bibinfo{year}{2007}).

\bibitem{2022YakuboPLOS}
\bibinfo{author}{Yakubo, K.} \& \bibinfo{author}{Fujiki, Y.}
\newblock \bibinfo{title}{A general model of hierarchical fractal scale-free
  networks}.
\newblock \emph{\bibinfo{journal}{PLoS ONE}} \textbf{\bibinfo{volume}{17}},
  \bibinfo{pages}{e0264589} (\bibinfo{year}{2022}).

\bibitem{2015KuangSciChina}
\bibinfo{author}{Kuang, L.}, \bibinfo{author}{Zheng, B.}, \bibinfo{author}{Li,
  D.}, \bibinfo{author}{Li, Y.} \& \bibinfo{author}{Sun, Y.}
\newblock \bibinfo{title}{A fractal and scale-free model of complex networks
  with hub attraction behaviors}.
\newblock \emph{\bibinfo{journal}{Sci. China Inf. Sci.}}
  \textbf{\bibinfo{volume}{58}}, \bibinfo{pages}{1--10} (\bibinfo{year}{2015}).

\bibitem{2022MolontayInProc}
\bibinfo{author}{Zakar-Poly{\'a}k, E.}, \bibinfo{author}{Nagy, M.} \&
  \bibinfo{author}{Molontay, R.}
\newblock \bibinfo{title}{Investigating the origins of fractality based on two
  novel fractal network models}.
\newblock In \bibinfo{editor}{Pacheco, D.} \& \bibinfo{editor}{et. al.} (eds.)
  \emph{\bibinfo{booktitle}{Complex Networks XIII}}, \bibinfo{pages}{43--54}
  (\bibinfo{publisher}{Springer International Publishing},
  \bibinfo{address}{Cham}, \bibinfo{year}{2022}).

\bibitem{2021BogunaNatPhys}
\bibinfo{author}{Bogu\~n\'a, M.} \emph{et~al.}
\newblock \bibinfo{title}{Network geometry}.
\newblock \emph{\bibinfo{journal}{Nat. Rev. Phys.}}
  \textbf{\bibinfo{volume}{3}}, \bibinfo{pages}{114--135}
  (\bibinfo{year}{2021}).

\bibitem{2007GallosPNAS}
\bibinfo{author}{Gallos, L.~K.}, \bibinfo{author}{Song, C.},
  \bibinfo{author}{Havlin, S.} \& \bibinfo{author}{Makse, H.~A.}
\newblock \bibinfo{title}{Scaling theory of transport in complex biological
  networks}.
\newblock \emph{\bibinfo{journal}{Proc. Natl. Acad. Sci. USA}}
  \textbf{\bibinfo{volume}{104}}, \bibinfo{pages}{7746--7751}
  (\bibinfo{year}{2007}).

\bibitem{2013GallosPLoS}
\bibinfo{author}{Gallos, L.~K.}, \bibinfo{author}{Potiguar, F.~Q.},
  \bibinfo{author}{Andrade~Jr, J.~S.} \& \bibinfo{author}{Makse, H.~A.}
\newblock \bibinfo{title}{Imdb network revisited: unveiling fractal and modular
  properties from a typical small-world network}.
\newblock \emph{\bibinfo{journal}{PLoS ONE}} \textbf{\bibinfo{volume}{8}},
  \bibinfo{pages}{e66443} (\bibinfo{year}{2013}).

\bibitem{1988bookFeder}
\bibinfo{author}{Feder, J.}
\newblock \emph{\bibinfo{title}{Fractals}} (\bibinfo{publisher}{Plenum Press},
  \bibinfo{address}{New York}, \bibinfo{year}{1988}).

\bibitem{2004bookMcComb}
\bibinfo{author}{McComb, W.~D.}
\newblock \emph{\bibinfo{title}{Renormalization Methods: A Guide for
  Beginners}} (\bibinfo{publisher}{Oxford University Press},
  \bibinfo{address}{New York}, \bibinfo{year}{2004}).

\bibitem{2009BundeEncyclopedia}
\bibinfo{author}{Bunde, A.} \& \bibinfo{author}{Havlin, S.}
\newblock \emph{\bibinfo{title}{Fractal geometry, a brief introduction to}}
  (\bibinfo{publisher}{Springer}, \bibinfo{address}{New York},
  \bibinfo{year}{2009}).

\bibitem{2019WeiPhysA}
\bibinfo{author}{Wei, B.} \& \bibinfo{author}{Deng, T.}
\newblock \bibinfo{title}{A cluster-growing dimension of complex networks: From
  the view of node closeness centrality}.
\newblock \emph{\bibinfo{journal}{Physica A}} \textbf{\bibinfo{volume}{522}},
  \bibinfo{pages}{80--87} (\bibinfo{year}{2019}).

\bibitem{1999AlbertNature}
\bibinfo{author}{Albert, R.}, \bibinfo{author}{Jeong, H.} \&
  \bibinfo{author}{Barab\'asi, A.-L.}
\newblock \bibinfo{title}{Diameter of the world wide web}.
\newblock \emph{\bibinfo{journal}{Nature}} \textbf{\bibinfo{volume}{401}},
  \bibinfo{pages}{130--131} (\bibinfo{year}{1999}).

\bibitem{DBLP1}
\bibinfo{author}{Tang, J.}, \bibinfo{author}{Fong, A. C.~M.},
  \bibinfo{author}{Wang, B.} \& \bibinfo{author}{Zhang, J.}
\newblock \bibinfo{title}{A unified probabilistic framework for name
  disambiguation in digital library}.
\newblock \emph{\bibinfo{journal}{IEEE Transactions on Knowledge and Data
  Engineering}} \textbf{\bibinfo{volume}{24}}, \bibinfo{pages}{975--987}
  (\bibinfo{year}{2012}).

\bibitem{2022FronczakSciRep}
\bibinfo{author}{Fronczak, A.}, \bibinfo{author}{Mrowinski, M.} \&
  \bibinfo{author}{Fronczak, P.}
\newblock \bibinfo{title}{Scientific success from the perspective of the
  strength of weak ties}.
\newblock \emph{\bibinfo{journal}{Sci. Rep.}} \textbf{\bibinfo{volume}{12}},
  \bibinfo{pages}{5074} (\bibinfo{year}{2022}).

\bibitem{2012GallosPNAS}
\bibinfo{author}{Gallos, L.~K.}, \bibinfo{author}{Makse, H.~A.} \&
  \bibinfo{author}{Sigman, M.}
\newblock \bibinfo{title}{A small world of weak ties provides optimal global
  integration of self-similar modules in functional brain networks}.
\newblock \emph{\bibinfo{journal}{Proc. Natl. Acad. Sci. USA}}
  \textbf{\bibinfo{volume}{109}}, \bibinfo{pages}{2825--2830}
  (\bibinfo{year}{2012}).

\bibitem{2014ReisNatPhys}
\bibinfo{author}{Reis, S. D.~S.} \emph{et~al.}
\newblock \bibinfo{title}{Avoiding catastrophic failure in correlated networks
  of networks}.
\newblock \emph{\bibinfo{journal}{Nat. Phys.}} \textbf{\bibinfo{volume}{10}},
  \bibinfo{pages}{762–767} (\bibinfo{year}{2014}).

\bibitem{1999BarabasiScience}
\bibinfo{author}{Barab\'asi, A.-L.} \& \bibinfo{author}{Albert, R.}
\newblock \bibinfo{title}{Emergence of scaling in random networks}.
\newblock \emph{\bibinfo{journal}{Science}} \textbf{\bibinfo{volume}{286}},
  \bibinfo{pages}{509–512} (\bibinfo{year}{1999}).

\bibitem{2007SongJStatMech}
\bibinfo{author}{Song, C.}, \bibinfo{author}{Gallos, L.~K.},
  \bibinfo{author}{Havlin, S.} \& \bibinfo{author}{Makse, H.~A.}
\newblock \bibinfo{title}{How to calculate the fractal dimension of a complex
  network: the box covering algorithm}.
\newblock \emph{\bibinfo{journal}{J. Stat. Mech.: Theory Exp.}}
  \textbf{\bibinfo{volume}{2007}}, \bibinfo{pages}{P03006}
  (\bibinfo{year}{2007}).

\bibitem{2008SerranoPRL}
\bibinfo{author}{Serrano, M.~A.}, \bibinfo{author}{Krioukov, D.} \&
  \bibinfo{author}{Bogu\~n\'a, M.}
\newblock \bibinfo{title}{Self-similarity of complex networks and hidden metric
  spaces}.
\newblock \emph{\bibinfo{journal}{Phys. Rev. Lett.}}
  \textbf{\bibinfo{volume}{100}}, \bibinfo{pages}{078701}
  (\bibinfo{year}{2008}).

\bibitem{2010KrioukovPRE}
\bibinfo{author}{Krioukov, D.}, \bibinfo{author}{Papadopoulos, F.},
  \bibinfo{author}{Kitsak, M.}, \bibinfo{author}{Vahdat, A.} \&
  \bibinfo{author}{Bogu\~n\'a, M.}
\newblock \bibinfo{title}{Hyperbolic geometry of complex networks}.
\newblock \emph{\bibinfo{journal}{Phys. Rev. E}} \textbf{\bibinfo{volume}{82}},
  \bibinfo{pages}{036106} (\bibinfo{year}{2010}).

\bibitem{2013BrockmannScience}
\bibinfo{author}{Brockmann, D.} \& \bibinfo{author}{Helbing, D.}
\newblock \bibinfo{title}{The hidden geometry of complex, network- driven
  contagion phenomena}.
\newblock \emph{\bibinfo{journal}{Science}} \textbf{\bibinfo{volume}{342}},
  \bibinfo{pages}{1337–1342} (\bibinfo{year}{2013}).

\bibitem{2015TaylorNatComm}
\bibinfo{author}{Dane~Taylor, D.} \emph{et~al.}
\newblock \bibinfo{title}{Topological data analysis of contagion maps for
  examining spreading processes on networks}.
\newblock \emph{\bibinfo{journal}{Nat. Commun.}} \textbf{\bibinfo{volume}{6}},
  \bibinfo{pages}{7723} (\bibinfo{year}{2015}).

\bibitem{2017DeDomenicoPRL}
\bibinfo{author}{De~Domenico, M.}
\newblock \bibinfo{title}{Diffusion geometry unravels the emergence of
  functional clusters in collective phenomena}.
\newblock \emph{\bibinfo{journal}{Phys. Rev. Lett.}}
  \textbf{\bibinfo{volume}{118}}, \bibinfo{pages}{168301}
  (\bibinfo{year}{2017}).

\bibitem{2021KovacsApplNetwSci}
\bibinfo{author}{Kov\'acs, P.~T.}, \bibinfo{author}{Nagy, M.} \&
  \bibinfo{author}{Molontay, R.}
\newblock \bibinfo{title}{Comparative analysis of box-covering algorithms for
  fractal networks}.
\newblock \emph{\bibinfo{journal}{Appl. Netw. Sci.}}
  \textbf{\bibinfo{volume}{6}}, \bibinfo{pages}{73} (\bibinfo{year}{2021}).

\bibitem{2012PapadopoulosNature}
\bibinfo{author}{Papadopoulos, F.}, \bibinfo{author}{Kitsak, M.},
  \bibinfo{author}{Serrano, M.~A.}, \bibinfo{author}{Bogu\~n\'a, M.} \&
  \bibinfo{author}{Krioukov, D.}
\newblock \bibinfo{title}{Popularity versus similarity in growing networks}.
\newblock \emph{\bibinfo{journal}{Nature}} \textbf{\bibinfo{volume}{489}},
  \bibinfo{pages}{537–540} (\bibinfo{year}{2012}).

\bibitem{2015ZuevSciRep}
\bibinfo{author}{Zuev, K.}, \bibinfo{author}{Bogu\~n\'a, M.},
  \bibinfo{author}{Bianconi, G.} \& \bibinfo{author}{Krioukov, D.}
\newblock \bibinfo{title}{Emergence of soft communities from geometric
  preferential attachment}.
\newblock \emph{\bibinfo{journal}{Sci. Rep.}} \textbf{\bibinfo{volume}{5}},
  \bibinfo{pages}{9421} (\bibinfo{year}{2015}).

\bibitem{2018GarciaPerezJStatPhys}
\bibinfo{author}{Garc\'ia-P\'erez, G.}, \bibinfo{author}{Serrano, M.~A.} \&
  \bibinfo{author}{Bogu\~n\'a, M.}
\newblock \bibinfo{title}{Soft communities in similarity space}.
\newblock \emph{\bibinfo{journal}{J. Stat. Phys.}}
  \textbf{\bibinfo{volume}{173}}, \bibinfo{pages}{775–782}
  (\bibinfo{year}{2018}).

\bibitem{2021KovacsSciRep}
\bibinfo{author}{Kov\'acs, B.}, \bibinfo{author}{Palla, G.},
  \bibinfo{author}{Der\'enyi, I.}, \bibinfo{author}{Farkas, I.} \&
  \bibinfo{author}{Vicsek, T.}
\newblock \bibinfo{title}{The inherent community structure of hyperbolic
  networks}.
\newblock \emph{\bibinfo{journal}{Sci. Rep.}} \textbf{\bibinfo{volume}{11}},
  \bibinfo{pages}{16050} (\bibinfo{year}{2021}).

\bibitem{2005PallaNature}
\bibinfo{author}{Palla, G.}, \bibinfo{author}{Der\'enyi, I.},
  \bibinfo{author}{Farkas, I.} \& \bibinfo{author}{Vicsek, T.}
\newblock \bibinfo{title}{Uncovering the overlapping community structure of
  complex networks in nature and society}.
\newblock \emph{\bibinfo{journal}{Nature}} \textbf{\bibinfo{volume}{435}},
  \bibinfo{pages}{814--818} (\bibinfo{year}{2005}).

\bibitem{2010FortunatoPhysRep}
\bibinfo{author}{Fortunato, S.}
\newblock \bibinfo{title}{Community detection in graphs}.
\newblock \emph{\bibinfo{journal}{Phys. Rep.}} \textbf{\bibinfo{volume}{486}},
  \bibinfo{pages}{75--174} (\bibinfo{year}{2010}).

\bibitem{2008bookBarrat}
\bibinfo{author}{Barrat, A.}, \bibinfo{author}{Barth\'elemy, M.} \&
  \bibinfo{author}{Vespignani, A.}
\newblock \emph{\bibinfo{title}{Dynamical Processes on Complex Networks}}
  (\bibinfo{publisher}{Cambridge University Press}, \bibinfo{year}{2008}).

\bibitem{2023FronczakSM}
\bibinfo{author}{Fronczak, A.} \emph{et~al.}
\newblock \bibinfo{title}{Supplementary materials for ''{S}caling theory of
  fractal complex networks: Bridging local self-similarity and global
  scale-invariance''}.
\newblock \bibinfo{howpublished}{Mendeley Data, V1, doi:10.17632/zn4xfhxym7.1}
  (\bibinfo{year}{2023}).

\bibitem{WWW}
\bibinfo{author}{Rossi, R.~A.} \& \bibinfo{author}{Ahmed, N.~K.}
\newblock \bibinfo{title}{The network data repository with interactive graph
  analytics and visualization}.
\newblock In \emph{\bibinfo{booktitle}{AAAI}} (\bibinfo{year}{2015}).
\newblock \urlprefix\url{https://networkrepository.com}.

\bibitem{DBLP2}
\bibinfo{author}{{DBLP Citation Network Dataset}}.
\newblock \bibinfo{howpublished}{\url{https://www.aminer.org/citation}}.
\newblock \bibinfo{note}{Accessed: 2022-08-30}.

\bibitem{BRAIN}
\bibinfo{howpublished}{\url{http://www-levich.engr.ccny.cuny.edu/~min/}}.
\newblock \bibinfo{note}{Accessed: 2022-01-30}.

\end{thebibliography}

\section*{Acknowledgements}

Research was funded by POB Cybersecurity and Data Science (AF, MS, KM, MJM) and POSTDOC PW programmes (PF, MŁ) of Warsaw University of Technology within the Excellence Initiative: Research University (IDUB). 

\section*{Author contributions statement}

A.F. designed and performed research, wrote the manuscript; P.F. performed research, wrote simulations and analysed data; M.S., K.M., M.Ł. and M.J.M. performed computer simulations and analysed data. All authors discussed the results and reviewed the manuscript.

\end{document}